\documentclass[twocolumn,prd,superscriptaddress,nofootinbib,preprintnumbers]{revtex4}
\usepackage{amsmath,amsfonts,amssymb}
\usepackage{graphicx}
\usepackage{dcolumn}
\usepackage[mathlines]{lineno}
\usepackage{float}
\usepackage{hyperref}
\usepackage[usenames,dvipsnames]{color}
\usepackage{etex}
\usepackage{booktabs}
\usepackage{adjustbox}

\newcommand\norm[1]{\lVert#1\rVert}

\begin{document}

\preprint{CTPU-PTC-23-52}

\title{Cosmological domain walls from the breaking of $\mathbf{S_4}$ flavor symmetry}
\author{Adil Jueid}
\affiliation{Particle Theory and Cosmology Group, Center for Theoretical Physics of the Universe, Institute for Basic Science (IBS), 34126 Daejeon, Republic of Korea}
\email{adiljueid@ibs.re.kr}
\author{Mohamed Amin Loualidi}
\affiliation{Department of physics, United Arab Emirates University, Al-Ain, UAE}
\email{ma.loualidi@uaeu.ac.ae}
\author{Salah Nasri}
\affiliation{Department of physics, United Arab Emirates University, Al-Ain, UAE}
\affiliation{The Abdus Salam International Centre for Theoretical Physics, Strada Costiera 11, I-34014, Trieste, Italy}
\email{snasri@uaeu.ac.ae, salah.nasri@cern.ch}
\author{Mohamed Amine Ouahid}
\affiliation{LPHE-Modeling and Simulations, Faculty of Science, Mohammed V University in Rabat, Morocco and Centre of Physics and Mathematics, CPM-Morocco}
\email{mohamedamine_ouahid@um5.ac.ma}

\begin{abstract}
In this work, we delve into the often-overlooked cosmological implications of spontaneous breaking of non-Abelian discrete groups, specifically focusing on the formation of domain walls in the case of $S_{4}$ flavor symmetry. In particular, we investigate three interesting breaking patterns of the $S_4$ group and study the structure of the domain walls in the broken phase for three possible residual symmetries. The presentation of domain walls in the case of multiple vacua is usually complicated, which therefore implies that most of the analyzes only approximate their presentation. Here, we propose a subtle way to represent the $S_{4}$ domain wall networks by presenting the vacua in each breaking pattern as vectors with their components corresponding to their coordinates in the flavon
space. Then, through the properties of the obtained vectors, we find that the domain wall networks can be represented by Platonic or Archimedean solids where the vertices represent the degenerate vacua while the edges correspond to the domain walls that separate them. As an illustration, we propose a type-II seesaw model that is based on the $S_{4}$ flavor symmetry, and study its phenomenological implications on the neutrino sector. To solve the domain wall problem within this toy model, we consider an approach based on high-dimensional effective operators induced by gravity that explicitly break the structure of the induced vacua favoring one vacuum over the others.
\end{abstract}


\maketitle
\section{Introduction}
\label{sec:intro}
In the past two decades, non-Abelian discrete symmetries have become more prominent in flavor model building. This interest grew particularly following the observation of large leptonic mixing angles from various neutrino oscillation experiments (see Refs. \cite{A1,A2,A3} for updated global fits). Non-Abelian discrete groups are increasingly employed to tackle the flavor problem which is connected to the lack of a mechanism within the Standard Model (SM) that explains the mass hierarchies of the different fermions and their mixing. Nevertheless, in order to achieve realistic predictions for the fermion masses and mixing angles at low energies, it is imperative to break the non-Abelian group due to the fact that the charged leptons and the three massive neutrinos are inherently distinct in the symmetric phase of the underlying flavor group. This breaking takes place when scalar fields called flavons, which are singlets under the SM gauge group, acquire {\it nonzero} vacuum expectation values (VEVs) along specific directions within the flavon space. On the other hand, the spontaneous symmetry breaking (SSB) of the discrete groups gives rise to multiple degenerate vacua separated by  surface-like topological defects referred to as domain walls (DWs) \cite{A4,A5}. The number of the vacua and DWs depends on the order of the broken group. Generally, these multiple vacua can be understood as the constituents of a topologically nontrivial vacuum manifold, defined by disconnected points in space with the same energy \cite{B3}. Once the field responsible for SSB stabilizes at one of these points ---no point is preferred over any other--- it cannot transition to any of the remaining points \cite{A6}. The regions between the degenerate ground states represent the DWs which, if proven to be stable, are disfavored by cosmological observations. To illustrate why this is an issue, notice that DWs are expected to enter a regime of dynamical scaling such that the number of defects is constant per Hubble horizon \cite{A7,A8,A9}. In this scaling regime, we have the property $H^{-1}\sim L\sim R$ where $H^{-1}$ is the Hubble radius, $L$ is the distance between two neighboring walls and $R$ is the wall curvature radius which is proportional to the cosmological expansion factor $a(t)$. The energy density of DWs in this regime scales as $a^{-1}$, decreasing more slowly than radiation or matter which scales as $a^{-4}$ or $a^{-3}$ respectively \cite{A4,A5,B3}. As a result, DWs could eventually dominate the Universe at small redshift which is disfavored by the current cosmological observations; this is known as the domain wall problem \cite{A4}. Indeed, Zel'dovich, Kobzarev and Okun pointed out that stable DWs would cause a fast expansion of the Universe dramatically affecting the formation of galaxies and reducing the production rates of light elements during primordial Nucleosynthesis \cite{A4}. Moreover, the presence of the walls in the current Universe would create unacceptable distortions in the cosmic microwave background (CMB) radiation that would violate the present limits on its homogeneity and isotropy \cite{A6}.\newline

Many approaches have been suggested to deal with the creation of DWs where the main idea is that they should either be unstable or remain subdominant until the present time. In particular, the energy scale associated with the SSB of the discrete groups should be low enough so that the energy density of the walls is a subdominant contribution to the total energy density of the Universe \cite{A10,D34}. This scale is also restrained to be smaller that $1$ MeV to prevent creating undesirable large anisotropies in the CMB \cite{A4}. Besides, the walls may exhibit instability if they manifest prior to the inflation era. However, below the inflationary scale, the most known solution to the DW problem was suggested by Zel'dovich et al. \cite{A4} where the DWs are unstable by assuming that the discrete symmetry is not exact. In other words, the introduction of explicit symmetry breaking terms create energy gaps between true and false vacua and in which case false vacua will fade before the walls dominate the energy density of the Universe (see also Refs. \cite{A5,B3,A11} for more details). The connection between the DW problem and flavor models that lead to it is usually overlooked in the literature. For instance, only the mechanism of DW creation and some of the solutions have been discussed in Refs. \cite{A12,A13,A14,A15,A16,B0} where they used $A_4$ or $D_4$ flavor symmetries. \newline

In this work, we investigate for the {\it first time} the formation of DWs from the SSB of the flavor group $S_{4}$. Here, we emphasize on three $S_{4}$ breaking patterns that have been established as phenomenologically plausible, {\it i.e.} $S_{4}\rightarrow Z_{2}\times Z_{2}$, $S_{4}\rightarrow Z_{3}$, and $S_{4}\rightarrow Z_{2}$. Since for each of these patterns $S_{4}$ is only partially broken, the number of degenerate vacua is contingent upon the order of the broken subgroup of $S_{4}$\footnote{In the context of $Z_{n} $ Abelian groups, the vacuum manifold is composed of $n$ degenerate vacua. The higher the order of the group, the greater the number of vacuum becomes leading to nontrivial DW networks.}. For the above $S_{4}$ breaking patterns, the broken subgroups are given by the non-Abelian groups $S_{3}$, $\Sigma (8)$ and $A_{4}$, respectively. Therefore, the transformations among the vacua are characterized by non-Abelian structures making the representation of the vacuum manifold even more complicated. Here, we propose a subtle way to represent the $S_{4}$ DW networks by using the properties of the flavon space which is defined as a vector space that can accommodate all the dimensions of the $S_{4}$ irreducible representations\footnote{This is a vector space of dimension six where the first three components are reserved for $S_{4}$ triplets, the next two for $S_{4}$ doublets and the last for $S_{4}$ singlets.}. After the SSB of $S_{4}$, the flavon field responsible for the breaking, say $\Omega$, can be represented by a vector where its components correspond to its coordinates in the vector space. The remaining vectors (vacua) are obtained easily by applying the various elements of the broken subgroups of $S_{4}$ on $\left\langle \Omega \right\rangle$. Then, through the properties of the obtained vectors, we find that the DW networks can be represented by the Platonic or Archimedean solids where the vertices represent the degenerate vacua while the edges represent the domain walls. To address the DW problem and its possible solution in the case of the $S_{4}$ flavor group, we propose a toy model based on type-II seesaw mechanism which we confront to the recent neutrino data. The problem of DWs in this toy model is solved by introducing effective operators generated by gravity and suppressed by powers of the Planck mass which simply means that the $S_4$ symmetry is not exact after all.

The rest of this paper is organized as follows. In Sec. \ref{sec2}, we start by identifying the $S_4$ breaking patterns consistent with the neutrino oscillation data. Then, we determine the breaking parts for each pattern, detailing properties crucial for DW formation. Finally, we offer a geometric description of DWs for each $S_4$ breaking pattern. In Sec. \ref{sec:Model}, after exploring some well-known solutions to the DW problem, we build a toy model with $S_4$ flavor symmetry and examine neutrino phenomenology as well as propose a possible solution to the DW problem. We also qualitatively discuss the gravitational waves (GWs) arising from DWs in this toy model. We conclude in Sec. \ref{concl}.
\section{Breaking of $S_4$ Symmetry and domain walls}
\label{sec2}
In this section, we first specify the different $S_{4}$ symmetry breaking patterns that are known to be phenomenologically consistent with the neutrino oscillation data, and we define some notations for the irreducible representations of $S_{4}$ as well as its subgroups. Afterwards, we focus on identifying the broken parts for each breaking pattern and describe their properties relevant for the creation of domain walls. Finally, we provide a geometric description of the DWs generated for each $S_{4}$ breaking pattern.
\subsection{Lepton residual symmetries from $S_4$}
The breaking of non-Abelian discrete groups $G_f$ is an essential part in the construction of flavor models to provide realistic predictions for lepton mixing angles. This breaking occurs when flavon fields acquire VEVs along specific directions in flavon space. Typically, the breaking of the flavor symmetry allows for the survival of different Abelian residual symmetries $G_{\rm res}$, which are subgroups of $G_f$. These residual symmetries give rise to distinct mixing patterns that can be explicitly derived by calculating the fermion mass matrices. Hence, the phenomenological viability of the fermion flavor structure is often assessed based on these surviving residual symmetries after the breaking of the underlying flavor group $G_f$. In this study, we examine the formation of DWs assuming $S_{4}$ as our flavor group, which breaks down into $G_{e}$ and $G_{\nu}$ associated to the charged lepton and the neutrino sectors, respectively, and assuming that neutrinos are Majorana particles.

\begin{figure}[!htb]
    \centering
    \includegraphics[width=0.95\linewidth]{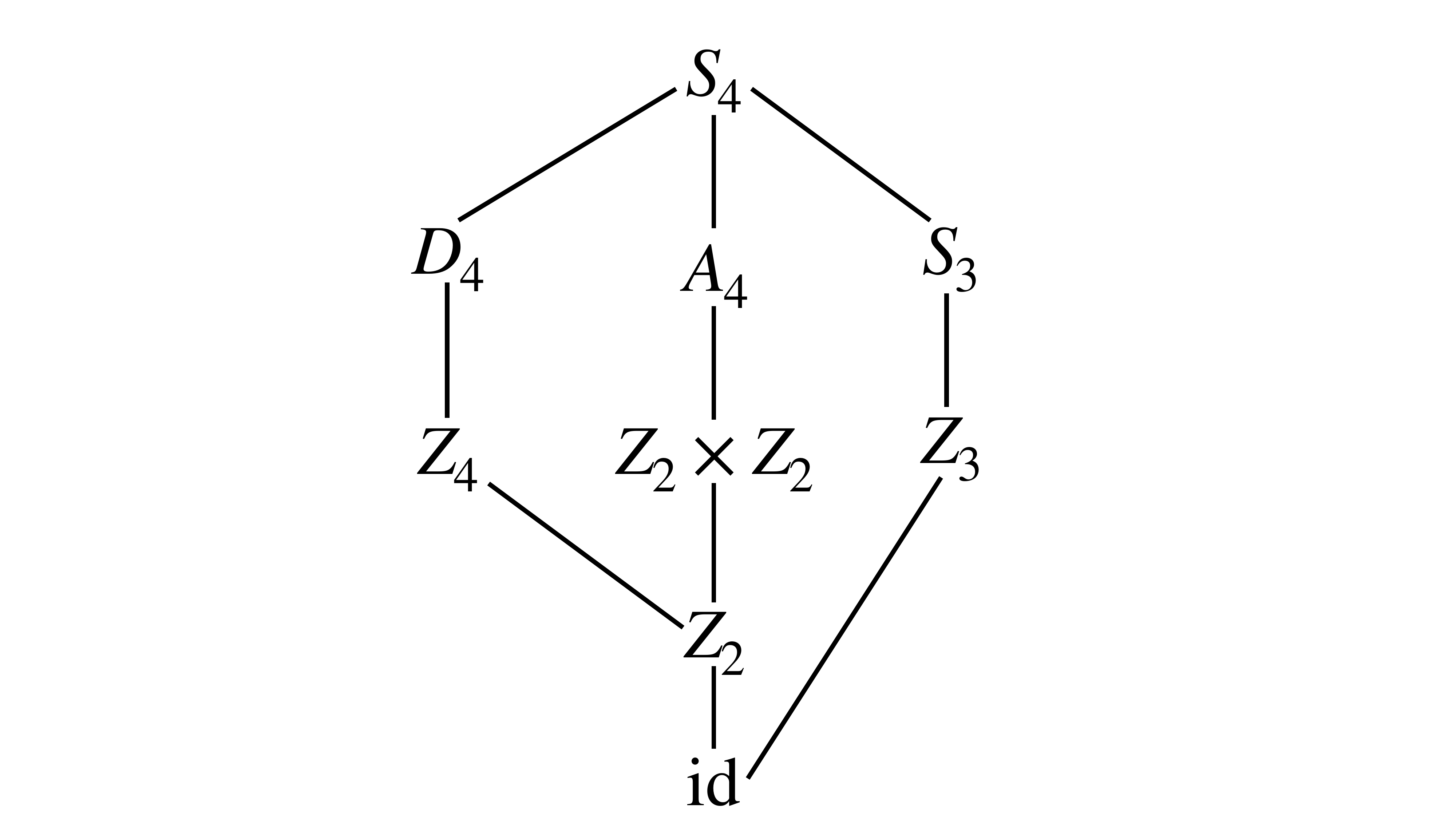}
    \caption{Symmetry breaking patterns of $S_4$ discrete symmetry.}
    \label{fig:S4:breaking}
\end{figure}

Depending on the flavon VEV alignment, the residual symmetry group
corresponding to any one of the $S_{4}$ subgroups is illustrated in Fig. \ref{fig:S4:breaking} where the residual symmetry group could be non-Abelian $\left\{A_{4},D_{4},S_{3}\right\} $ or Abelian $\left\{ Z_{4},Z_{2}\times
Z_{2},Z_{3}\right\}$.
The breaking of $S_{4}$ down to one of its non-Abelian subgroups is excluded by current data since it leads to two degenerate neutrino mass states. For the case of Abelian subgroups, it is convenient to use the fact that $S_{4}$ is isomorphic to $G_{f}^{\prime} =(Z_{2}\times Z_{2})\rtimes \left(Z_{3}\rtimes Z_{2}\right) $, where $(Z_{2}\times Z_{2})$ refers to the Klein four-group denoted by $K_{4}$ in what follows, while $\left( Z_{3}\rtimes Z_{2}\right) $ is isomorphic to the smallest non-Abelian group; the symmetric group $S_{3}$. Consequently, we can perceive $S_{4}$ as the semi-direct product of $K_{4}$ and $S_{3}$, denoted as $S_{4}\cong K_{4}\rtimes S_{3}$. This group contains $24$ elements generated by three generators denoted by ${\cal S}$, ${\cal T}$ and ${\cal U}$. It is noteworthy, though, that only two generators are required to define $S_4$ \cite{D1}. However, in order to highlight the relationship between the alternating group $A_{4}$ and $S_{4}$, it is advantageous to choose the set of generators as $\mathcal{S}$, $\mathcal{T}$ and $\mathcal{U}$. By doing so, it becomes evident that $\mathcal{S}$ and $\mathcal{T}$ alone can generate the group $A_{4}$ \cite{D2}. Similarly, the two generators $\mathcal{T}$ and $\mathcal{U}$ alone can generate the group $S_{3}$ \cite{D3}. The $S_{4}$ group consists of five conjugacy classes and therefore contains five irreducible representations. These representations
include two singlets: $\mathbf{1}_{(1,1,1)}$ (trivial) and $\mathbf{1}_{(1,1,-1)}^{\prime }$, one doublet $\mathbf{2}_{(2,-1,0)}$ and two triplets 
$\mathbf{3}_{(-1,0,-1)}$ and $\mathbf{3}_{\left( -1,0,1\right) }^{\prime }$ with the indices referring to the characters of the generators $\mathcal{S}$, $\mathcal{T}$ and $\mathcal{U}$ respectively; see the Tab. \ref{tab:characters}. In addition, $S_{4}$ has $20$ Abelian subgroups, which consist of nine $Z_{2}$ subgroups, four $Z_{3}$ subgroups, three $Z_{4}$ subgroups, and four Klein-four subgroups ($K_{4}$). These symmetries can be expressed in terms of the $S_{4}$ generators as provided in Eq. \ref{eq:S4:subgroups} of the appendix.\newline

Before going through the details of the $S_{4}$ breaking schemes, we would like to make the following comments relevant to the analysis of DWs formed under various  $S_{4}$ breaking patterns:
\begin{itemize}
\item In this study, we adopt the representation matrices for the $S_{4}$ generators in the basis where $\mathcal{T}$ is diagonal (see Eq. \ref{t2}) in the appendix. This particular basis is commonly used in the literature since it leads to a diagonal charged lepton mass matrix which remains invariant under $G_{e}=Z_{3}^{\mathcal{T}}$ generated by $\mathcal{T}$. Moreover, $G_{e}=Z_{4}$ is ruled out since it leads to values of the solar mixing angle outside $3\sigma$ range reported on by the authors of Ref. \cite{D4} while $G_{\nu}$ is either $K_{4}$ or $Z_{2}$ \cite{D5}. Consequently, $Z_{4}$ is discarded as a phenomenologically viable residual group.

\item One of our main interests is to search, for {\it the first time}, for graphical representations of the DW networks created during SSB of the $S_{4}$ group. Thus, we discuss the number of degenerate vacua extracted from the broken subgroups within each $S_4$ breaking pattern. In order to clarify this point from the beginning, consider the scenario of SSB of $S_{4}$ into one of its four $Z_{3}$ subgroups. In this case, the broken part corresponds to $K_{4} \rtimes Z_{2}$, which is an order-8 structure that also represents the number of degenerate vacua, regardless of the specific $Z_{3}$'s, $K_{4}$'s and $Z_{2}$'s subgroups of $S_{4}$ involved in this breaking pattern. For this reason, choosing one of the four $Z_{3}$'s as the residual symmetry makes no difference for the study of the DWs. The same reasoning applies if the breaking pattern is into one of the nine $Z_{2}$ subgroups and/or the four $K_{4}$ subgroups of $S_{4}$.

\item The SSB of $S_{4}$ symmetry breaking can occur via a nontrivial $S_{4}$-singlet, an $S_{4}$-doublet or an $S_{4}$-triplet whose VEVs preserve one of the subgroups of $S_{4}$. To simplify our discussion, we focus solely on the scenario where a flavon field is designated to transform as an $S_{4}$-triplet. However, we will provide additional comments whenever is necessary for the other cases.

\item The basis mentioned in the first comment is particularly interesting and useful for model building. However, in order to graphically represent the components that have undergone SSB, and thus to investigate the properties of DWs, we need to express the obtained degenerate vacua in a three-dimensional real basis. In this picture, $\mathcal{T}$ can be represented by a diagonal $3\times 3$ matrix with eigenvalues $1, \omega$, and $\omega^2$ and its typical eigenvectors $x_{q}^{\prime }$ are respectively given by
\begin{eqnarray}
x_{1}^{\prime } &=& x_{1}+x_{2}+x_{3}, \nonumber \\ 
x_{2}^{\prime } &=& x_{1}+\omega^{2}x_{2}+\omega x_{3}, \nonumber \\ x_{3}^{\prime } &=& x_{1}+\omega x_{2}+\omega^{2}x_{3},
\end{eqnarray}
with $x_1$, $x_2$ and $x_3$ being the components of a complex vector space that defines the system basis for the triplet representation of $S_4$.
\end{itemize}
\subsection{Symmetry breaking patterns of  $S_4$}
In view of the above, we will examine three breaking patterns of $S_{4}$ that are phenomenologically viable. Our study does not require the identification of the residual group between $G_e$ and $G_\nu$. Therefore, we do not specify here the sector within which $S_4$ breaking takes place but only later when we introduce the toy model that requires the definition of the scale at which $S_4$ is broken. Refs. \cite{D4, D6} explore all the possible combinations of the pairs $G_e, G_\nu$. Let us now establish the $S_{4}$ breaking patterns within the basis provided in equation \ref{t2}:\newline

\emph{(a)} ${\bf S_{4}\longrightarrow K_{4}}$: Assuming that the residual $K_{4}$ symmetry is generated by $\{\mathcal{S},\mathcal{U}\}$, this breaking can be realized by the $S_{4}$ triplet $\Phi^{\prime}\equiv \mathbf{3}_{(-1,0,1)}^{\prime }$ acquiring a VEV along the direction $\left\langle \Phi ^{\prime }\right\rangle =\upsilon _{\Phi }(1,1,1)^{T}$. Here, we should be careful about the generator of the third $Z_{2}$ group in $G_{f}^{\prime }$ ---besides\footnote{The superscripts denote the generators for each $Z_{2}$ cyclic group.} $K_{4}\cong (Z_{2}^{\mathcal{S}}\times Z_{2}^{\mathcal{U}})$--- given that its generator should be broken in order for $K_{4}$ to be the only residual group. For this reason, we consider that the third cyclic group $Z_{2}\subset G_{f}^{\prime}$ is generated by $\mathcal{T}^{2}\mathcal{ST}$ which is the generator of one of the nine $Z_{2}$ subgroups of $S_{4}$ (see Eq. \ref{eq:S4:subgroups}). By exhibiting this breaking using the matrix representations of the generators of $S_{4}$ shown in Eq. \ref{t2}, we find 
\begin{eqnarray}
    {\cal S} \left\langle \Phi^\prime \right\rangle &=& \left\langle \Phi^\prime \right\rangle, \quad 
    {\cal U} \left\langle \Phi^\prime \right\rangle = \left\langle \Phi^\prime \right\rangle, \nonumber \\ 
    {\cal T} \left\langle \Phi^\prime \right\rangle &\neq&  \left\langle \Phi^\prime \right\rangle, \quad 
    {\cal T}^2 {\cal S T} \left\langle \Phi^\prime \right\rangle \neq \left\langle \Phi^\prime \right\rangle.
    \label{z4}
\end{eqnarray}    
It is clear from this equation that the only preserved symmetry is $K_{4}\cong (Z_{2}^{\mathcal{S}}\times Z_{2}^{\mathcal{U}})$ while the broken part is given by the symmetric group $S_{3}\cong Z_{3}^{\mathcal{T}}\rtimes Z_{2}^{\mathcal{T}^{2}\mathcal{ST}}$. In the case of the $S_{4}$ doublet $\mathbf{2}_{(2,-1,0)}$, it is not possible to exclusively decompose $S_{4}$ into $K_{4}^{\{\mathcal{S}, \mathcal{U}\}}$. This is due to the fact that $Z_{2}^{\mathcal{T}^{2} \mathcal{ST}}$ is always conserved, as $\mathcal{T}^{2}\mathcal{ST}$ essentially represents a two-dimensional identity matrix. Consequently, regardless of the direction of any VEV, its application will yield an identical VEV direction. On the other hand, it is not possible to use the nontrivial singlet $\mathbf{1}_{(1,1,-1)}^{\prime }$ for this breaking because it transforms oppositely under the $\mathcal{S}$ and $\mathcal{U}$ generators (see the Appendix for more details). It is important to emphasize that all the details regarding the VEV alignments described above depend on the choice of generators for all the groups involved in $G_{f}^{\prime}$. For example, if we take a different $Z_{3}$ group in the breaking pattern $S_{4}\longrightarrow Z_{3}$ than $Z_{3}^{\mathcal{T}}$\footnote{The other choices for the breaking pattern $S_4 \to Z_3$ are $Z_{3}^{\mathcal{ST}}$ $,$ $Z_{3}^{\mathcal{TS}}$ and $Z_{3}^{\mathcal{STS}}$.} we may need a different VEV structure for the flavon triplet than the one chosen above to realize this breaking. Therefore, for different choices of the $S_{4}$ subgroups we may end up with different scalar sectors. However, this is completely model independent when it comes to the geometrical interpretation of the DWs created during each breaking pattern as we mentioned in the second comment above.\newline

\emph{(b)}  ${\bf S_{4}\longrightarrow Z_{3}}$: Let us
take the isomorphic group to be $G_{f}^{\prime }\simeq (Z_{2}^{\mathcal{S}}\times Z_{2}^{\mathcal{U}})\rtimes \left( Z_{3}^{\mathcal{T}}\rtimes Z_{2}^{\mathcal{TST}^{2}}\right)$, which means that the residual symmetry in this case is given by $Z_{3}^{\mathcal{T}}$. As a result, this breaking can be realized by the $S_{4}$ triplet $\Phi \equiv \mathbf{3}_{(-1,0,-1)}$ acquiring a VEV along the direction $\left\langle \Phi \right\rangle=\upsilon_{\Phi}(1,0,0)^{T}$ \cite{D7}. By exhibiting this breaking using the matrix representations of the generators of $S_{4}$ in Eq. (\ref{t2}), we obtain
\begin{eqnarray}
{\cal S} \left\langle \Phi \right\rangle &\neq& \left\langle \Phi \right\rangle,  \quad {\cal U} \left\langle \Phi \right\rangle \neq \left\langle \Phi \right\rangle, \nonumber \\ 
{\cal T} \left\langle \Phi \right\rangle &=& \left\langle \Phi \right\rangle, \quad {\cal T} {\cal S T}^2 \left\langle \Phi \right\rangle \neq \left\langle \Phi \right\rangle.
\label{z3}
\end{eqnarray} 
It is clear from this equation that the only preserved symmetry is $Z_{3}^{\mathcal{T}}$ while the broken part is given by $\Sigma \left( 8\right)
\equiv \left( Z_{2}^{\mathcal{S}}\times Z_{2}^{\mathcal{U}}\right) \rtimes
Z_{2}^{\mathcal{TST}^{2}}$ which is isomorphic to the dihedral $D_{4}$ group \cite{D8}. Notice by the way that if we use the other $S_{4}$\ triplet $\mathbf{3}_{\left( -1,0,1\right) }^{\prime }$\ with the same VEV structure, the $S_{4}$\ group will break down to $Z_{2}^{\mathcal{U}}\rtimes Z_{3}^{%
\mathcal{T}}$\ which is isomorphic to the symmetric $S_{3}$\ group. Consequently, the breaking pattern $S_{4}\longrightarrow Z_{3}^{\mathcal{T}}$ is not achieved with the triplet $\mathbf{3}_{\left( -1,0,1\right)}^{\prime }$ unless a different VEV structure is chosen. We must stress out that the $S_4$ triplet $\Phi \equiv \mathbf{3}_{(-1,0,-1)}$ have to acquire a VEV along the direction $\left\langle \Phi \right\rangle =\upsilon _{\Phi }(-1,2,2)^{T}$ if we need to realize the breaking pattern $S_4 \to Z_3^{\cal T}$ if the residual group is represented by $Z_3^{\cal STS}$. For the $S_{4}$ doublet $\mathbf{2}_{(2,-1,0)}$, there are no trivial VEV structures that allows the preservation of the $\mathcal{T}$ or $\mathcal{STS}$\ generators, while the concluding remark for the $S_{4}$ singlet $\mathbf{%
1}_{(1,1,-1)}^{\prime }$ in the previous breaking scheme holds as well for the $S_{4}$ breaking down to $Z_{3}$.\newline

$\emph{(c)}$  ${\bf S_{4}\longrightarrow Z_{2}}$: let us denote the three $Z_{2}$ replicas in the isomorphic group $G_{f}^{\prime}\simeq (Z_{2}\times Z_{2})\rtimes \left( Z_{3}\rtimes Z_{2}\right) $ by $Z_{2}^{\mathcal{U}}$, $Z_{2}^{\mathcal{S}}$ and $Z_{2}^{\mathcal{SU}}$ where we take the latter as the residual symmetry associated with the semi-direct product $\left( Z_{3}\rtimes Z_{2}\right) $. Assuming that this breaking is induced by the $S_{4}$ triplet $\Phi \equiv \mathbf{3}_{(-1,0,-1)}$, the VEV alignment in this case is given by\textbf{\ }$\left\langle \Phi
\right\rangle =\upsilon _{\Phi }(2,-1,-1)^{T}$ as shown in Ref. \cite{D7}. This can be easily exhibited using the matrix representations of the generators of $S_{4}$, given in Eq. \ref{t2}, as follows 
\begin{eqnarray}
    {\cal S} \left\langle \Phi \right\rangle &\neq& \left\langle \Phi \right\rangle,  \quad {\cal U} \left\langle \Phi \right\rangle \neq \left\langle \Phi \right\rangle, \nonumber \\
    {\cal T} \left\langle \Phi \right\rangle &\neq& \left\langle \Phi \right\rangle, \quad {\cal S} {\cal U} \left\langle \Phi \right\rangle = \left\langle \Phi \right\rangle.
    \label{z2}
\end{eqnarray}    
It is clear from this equation that the only preserved symmetry is $Z_{2}^{\mathcal{SU}}$ while the broken part is given by $Z_{2}^{\mathcal{U}}\times Z_{2}^{\mathcal{S}}\rtimes Z_{3}^{\mathcal{T}}$ which is isomorphic to the alternating $A_{4}$ group. It is worth noticing that the use of the other $%
S_{4}$ triplet $\mathbf{3}_{\left( -1,0,1\right) }^{\prime }$ while maintaining the same VEV structure leads to the breakdown of the $S_{4}$ group into its $Z_{2}$ subgroup generated by $\mathcal{U}$ where it is straightforward to verify that $\mathcal{U}\left\langle \Phi \right\rangle=\left\langle \Phi \right\rangle $. On the other hand, thinking of $Z_{2}^{%
\mathcal{SU}}$ as one of the $Z_{2}$ groups in the Klein four group leads to the broken part $Z_{2}^{\mathcal{U}}\rtimes Z_{2}^{\mathcal{S}}\rtimes Z_{3}^{\mathcal{T}}$ which is a different group compared to $Z_{2}^{\mathcal{U}}\times Z_{2}^{\mathcal{S}}\rtimes Z_{3}^{\mathcal{T}}$. However, both of these groups have the same order, thereby resulting in an equal number of degenerate vacua, as mentioned in the second comment provided above. We will now take a brief look at alternative representations of flavons. If we examine the breaking pattern $S_{4}\longrightarrow Z_{2}$ via an $S_{4}$ doublet $\mathbf{2}_{(2,-1,0)}$, the residual $Z_{2}$ symmetry generated by $\mathcal{SU}$ is no longer a viable choice. This is because $\mathcal{SU}$ and $\mathcal{U}$ transform in the same manner, leading to the same outcome when applied to $\left\langle \Phi \right\rangle $. However, if we designate the residual symmetry as $Z_{2}^{\mathcal{S}}$, we can achieve this breaking
pattern by adopting a VEV configuration as $\left\langle \Phi \right\rangle = \upsilon _{\Phi }(1,0)^{T}$. Another possibility is to opt for different $Z_{2}\in G_{f}^{\prime }$ subgroups among the nine available in $S_{4}$. On the other hand, it is not possible to use the nontrivial singlet $\mathbf{1}_{(1,1,-1)}^{\prime }$ for this breaking because it transforms in the same manner under $\mathcal{S}$ and $\mathcal{T}$ generators and thus the $\mathcal{T}$ generator will be always preserved together with $\mathcal{S}$. As a result, the nontrivial singlet will inevitably cause the breakdown of $S_{4}$ to a subgroup other than $Z_{2}$.
\subsection{DWs configuration}
\subsubsection{Introduction}
To understand how DWs are positioned in flavon space, it is advantageous to look for concrete visual representations. These
representations aid in illustrating the outcomes of symmetry breaking, ultimately facilitating the understanding of the connections among distinct regions that interpolate between the various $\Phi_{a}$-vacua. In particular, we are going to show that the use of networks or graphical representations to present DWs arising from the SSB of the $S_{4}$ group can be useful to capture the properties of these topological defect, such as their boundaries, intersections, and connectivity. Moreover, these graphical representations may also be interpreted as quiver diagrams in the flavon space; denoted in what follows by $\zeta$.

To establish a comprehensive framework for the $S_{4}$ group's flavon space, our approach involves envisioning a six-dimensional vector space $\mathbf{V}$ capable of accommodating the various dimensions associated with $S_{4}$ irreducible representations. Specifically, we represent this vector as ${\bf V} = \left(x_1, x_2, x_3, x_4, x_5, x_6\right) \in \mathbb{C}$. As mentioned in the previous section, our primary focus lies on $S_{4} $ triplets, leading us to conceive the flavon triplet $\mathbf{\Phi }$ in terms of the vector space $\mathbf{V}$ components as\footnote{A more general scenario would be to consider a collective flavon field $%
\mathbf{\digamma }\mathbf{\sim }\digamma _{k}x_{k}$ where $\mathbf{\digamma }%
=(\mathbf{\Phi },\mathbf{\varphi },\mathbf{\sigma })$ with $\mathbf{\varphi }%
=\varphi _{j}x_{j}, j=4,5$ being a flavon doublet while $\mathbf{\sigma }=\sigma x_{6}$ is a flavon singlet. As a result, we can think of these $%
x_{i}^{\prime }$s as the foundation of a vector basis system in a flavon space $\zeta $\ where the first three directions ($x_{1},x_{2},x_{3}$) correspond to the $S_{4}$- triplets, the subsequent two directions ($x_{4},x_{5}$) correspond to $S_{4}$-doublets and the last direction $x_{6}$ corresponds to $S_{4}$-singlets.}
\begin{equation}
\mathbf{\Phi }\sim \Phi _{i}x_{i}\quad ,\quad i=1,2,3
\end{equation}%
On the other hand, in order to use real graphical representation for DWs, we need to find a real representation for the flavon fields which are in general complex fields. We can approach this by considering the splitting of the flavon fields as well as the vector space $\mathbf{V}$ into real and imaginary parts. For the flavon triplets, we have 
\begin{equation}
\Phi_i = \mathbf{\Re}\left( \Phi_{i}\right) + i\mathbf{%
\Im}\left( \Phi_{i}\right), \quad \mathbf{V}_{i} = \mathbf{\Re}\left( U_{i}\right) \mathbf{+}i\mathbf{\Im}%
\left( R_{i}\right), i=1,2,3,
\end{equation} 
where $U_{i}$ and $R_{i}$ are real $3$D vectors that can be understood as the constituents of $x_{i}$ as $x_{1}=(U_{1},R_{1})^{T},...,x_{3}=(U_{3},R_{3})^{T}$. With this real representation, we can think of the complex $3$D expansion $\mathbf{\Phi }%
\mathbf{\sim }\Phi _{i}x_{i}$ as a real $6$ dimensional vector like
\begin{equation}
\mathbf{\Phi \sim }\sum_{i}\left[ \mathbf{\Re}\left( \mathbf{\Phi }_{i}\right) U_{i}+\mathbf{\Im}\left( \mathbf{\Phi }_{i}\right) R_{i}\right] \in \mathbb{R}
\end{equation}
For illustration, we demonstrated in the previous section that the $S_{4}\rightarrow K_{4}$ symmetry breaking pattern can be achieved through a flavon triplet $\Phi^{\prime }$ which attains a VEV along the direction $\left\langle \Phi ^{\prime }\right\rangle =\upsilon _{\Phi ^{\prime}}(1,1,1)^{T}$. This particular VEV can be expressed in the real six-dimensional basis as 
\begin{equation}
\left\langle \Phi ^{\prime }\right\rangle =\upsilon _{\Phi }(1,0,1,0,1,0)^{T}
\label{rv1}
\end{equation}%
Expanding upon this, there exists another VEV direction involving complex entries that corresponds to the $S_{4}\rightarrow K_{4}$ breaking pattern. We label this direction as $\left\langle \Phi ^{\prime \prime
}\right\rangle =\upsilon _{\Phi ^{\prime \prime }}(1,\omega ,\omega^{2})^{T} $ where $\omega =-1/2+i\sqrt{3}/2$ \cite{D7}. In this case, $\left\langle \Phi ^{\prime \prime }\right\rangle $ is represented in the real basis as 
\begin{equation}
\left\langle \Phi ^{\prime \prime }\right\rangle =\upsilon _{\Phi ^{\prime \prime }}(1,0,-1/2,\sqrt{3}/2,-1/2,-\sqrt{3}/2)^{T}  \label{rv2}
\end{equation}%
From a geometric perspective, the VEV $\left\langle \Phi ^{\prime
}\right\rangle $ responsible for the breaking can be represented by a vector situated in one of the real directions within the vector space $\left(U_{k},R_{k}\right) $. The components of this vector correspond to its coordinates within that space. Then, the transformations applied to $\left\langle \Phi ^{\prime }\right\rangle $ by the various elements of the $S_{4}$ group create a polygon in the space defined by $\left(U_k, R_k\right)$. 
The number of vertices in the polygon is equal to the number of degenerate vacua resulting from the SSB of the $S_{4}$ group. For example, when the $S_{4}$ group is completely broken, the VEV responsible for this breaking, say $\left\langle \Omega \right\rangle $, splits into $24$ vacua with the same energy $\left\langle \Omega \right\rangle _{i}\equiv \epsilon _{i}$ with $i=1,...,24$. Among these vacua, one is just the VEV used to break $S_{4}$; say $\epsilon _{1}=\left\langle \Omega \right\rangle $. The remaining 23 vacua emerge through the application of the $S_{4}$ group elements upon $\epsilon_{1}$. These transformations generate a polygon with the following 24 vertices
\begin{widetext}
\begin{eqnarray}
\epsilon _{1} &=&\left\langle \Omega \right\rangle,~\epsilon _{2}=\mathcal{S}\epsilon_{1}~,~\epsilon _{3}=\mathcal{TST}^{2}\epsilon _{1}~,~\epsilon
_{4}=\mathcal{T}^{2}\mathcal{ST}\epsilon _{1}~,~\epsilon _{5}=\mathcal{U}%
\epsilon _{1}~,~\epsilon _{6}=\mathcal{TU}\epsilon _{1}  \notag \\
\epsilon _{7} &=&\mathcal{SU}\epsilon _{1}~,~\epsilon _{8}=\mathcal{T}^{2}%
\mathcal{U}\epsilon _{1}~,~\epsilon _{9}=\mathcal{STSU}\epsilon
_{1}~,~\epsilon _{10}=\mathcal{ST}^{2}\mathcal{SU}\epsilon _{1}~,~\epsilon
_{11}=\mathcal{T}\epsilon _{1}~,~\epsilon _{12}=\mathcal{ST}\epsilon _{1} 
\notag \\
\epsilon _{13} &=&\mathcal{TS}\epsilon _{1}~,~\epsilon _{14}=\mathcal{STS}%
\epsilon _{1}~,~\epsilon _{15}=\mathcal{T}^{2}\epsilon _{1}~,~\epsilon _{16}=%
\mathcal{ST}^{2}\epsilon _{1}~,~\epsilon _{17}=\mathcal{T}^{2}\mathcal{S}%
\epsilon _{1}~,~\epsilon _{18}=\mathcal{ST}^{2}\mathcal{S}\epsilon _{1} \\
\epsilon _{19} &=&\mathcal{STU}\epsilon _{1}~,~\epsilon _{20}=\mathcal{TSU}%
\epsilon _{1}~,~\epsilon _{21}=\mathcal{T}^{2}\mathcal{SU}\epsilon
_{1}~,~\epsilon _{22}=\mathcal{ST}^{2}\mathcal{U}\epsilon _{1}~,~\epsilon
_{23}=\mathcal{TST}^{2}\mathcal{U}\epsilon _{1}~,~\epsilon _{24}=\mathcal{T}%
^{2}\mathcal{STU}\epsilon _{1}  \notag,
\end{eqnarray}%
\end{widetext}
where the elements of $S_{4}$ are provided in the appendix. The obtained polygonal graph can be interpreted as a DW quiver observed from the flavon space $\zeta $, where $S_{4}$ is completely broken. On the other hand, we mentioned in the previous section that the models that are phenomenologically viable for explaining the observed flavor structure correspond to partial breaking of $S_{4}$ down to one of its subgroups: $%
Z_{2}$, $Z_{3}$, or $K_{4}$. Therefore, the broken parts in each of these breaking scenarios will yield a reduced number of vertices compared to the scenario where $S_{4}$ is completely broken. In what follows, we will examine each case independently, with the aim of describing the DW properties by identifying the polygonal graphs associated with each breaking pattern.
\subsubsection{DWs for $S_{4}\rightarrow K_{4}$ breaking}
\begin{figure}[!t]
\centering
\includegraphics[width=0.8\linewidth]{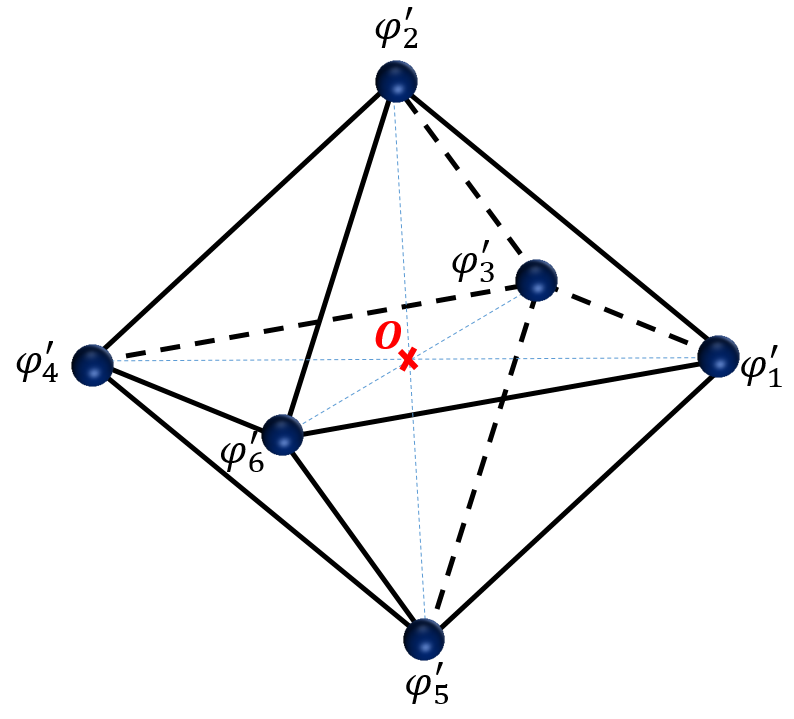}
\caption{A regular Octahedron viewed from the flavon space $\protect\zeta $. The six vertices represent the six degenerate vacua, while the edges that connect them stand for domain walls. }
\label{fig2}
\end{figure}

As outlined above, the breaking pattern $S_{4}\rightarrow K_{4}$ can be realized by the flavon triplet\ $\Phi ^{\prime }\equiv \mathbf{3}_{(-1,0,1)}^{\prime }$ acquiring a VEV along the direction $\left\langle \Phi ^{\prime }\right\rangle =\upsilon _{\Phi }(1,1,1)^{T}$. The Klein four group we have chosen for this study is $K_{4}\cong Z_{2}^{\mathcal{S}}\times
Z_{2}^{\mathcal{U}}$ and thus, this breaking can be expressed using $G_{f}^{\prime }$ as follows
\begin{equation}
Z_{2}^{\mathcal{S}}\times Z_{2}^{\mathcal{U}}\rtimes Z_{3}^{\mathcal{T}}\rtimes Z_{2}^{\mathcal{T}^{2}\mathcal{ST}}\overset{\left\langle \Phi
^{\prime }\right\rangle }{\longrightarrow }K_{4},
\end{equation}
where the broken part is given by $Z_{3}^{\mathcal{T}}$ $\rtimes Z_{2}^{\mathcal{T}^{2}\mathcal{ST}}$ which is isomorphic to the symmetric group $S_{3}$. The order of $S_{3}$ is six which means that the initial $S_{4}$ invariant vacuum gets now split into six vacua with the same energy $\left\langle \Phi \right\rangle _{i}\equiv \varphi _{i}$ with $i=1,...,6$.
These six vacua are positioned within the flavon space $\zeta$ and collectively establish the vertices of a Platonic solid known as a regular Octahedron as depicted in figure \ref{fig2}. One of the vertices defining the Octahedron is the $K_{4}$- invariant vacuum used for the $S_{4}\rightarrow K_{4}$ breaking; $\varphi _{1}^{\prime }=\left\langle \Phi^{\prime }\right\rangle $. To derive the remaining five $\varphi_{i}^{\prime }$\ vacua, we act upon $\varphi _{1}^{\prime }$\ with the generators of $Z_{3}^{\mathcal{T}}\rtimes Z_{2}^{\mathcal{T}^{2}\mathcal{ST}}
$ as follows
\begin{eqnarray}
\varphi_{1}^{\prime} &=& \left\langle \Phi^{\prime }\right\rangle, \quad \varphi_{2}^{\prime}=\mathcal{T}\varphi_{1}^{\prime}, \nonumber \\ 
\varphi_{3}^{\prime} &=& \mathcal{T}^{2}\varphi_{1}^{\prime }, 
\quad \varphi_{4}^{\prime} = \mathcal{T}^{2}\mathcal{ST}\varphi_{1}^{\prime }, \\ \varphi_{5}^{\prime} &=& \mathcal{T}\left(\mathcal{T}^{2}\mathcal{ST}\right) \varphi_{1}^{\prime}, \quad \varphi_{6}^{\prime} = \mathcal{T}^{2}\left( \mathcal{T}^{2}\mathcal{ST}\right) \varphi_{1}^{\prime}.  \nonumber
\end{eqnarray}
In the complex three-dimensional space $\mathbb{C}^{3}$, these transformations lead to
\begin{widetext}
\begin{equation}
\varphi_{1,4}^{\prime }=\pm \upsilon_{\Phi^{\prime}}\left( 
\begin{array}{c}
1 \\ 
1 \\ 
1
\end{array}
\right), \quad \varphi_{2,5}^{\prime}=\pm \upsilon_{\Phi^{\prime }}\left( 
\begin{array}{c}
1 \\ 
\omega^{2} \\ 
\omega 
\end{array}
\right), \quad \varphi _{3,6}^{\prime}=\pm \upsilon_{\Phi ^{\prime}}\left( 
\begin{array}{c}
1 \\ 
\omega  \\ 
\omega ^{2}%
\end{array}%
\right) 
\end{equation}  
\end{widetext}
In the real vector basis, we have to use real variables, so the dimensions of the above vectors has to be doubled $\mathbb{C}^{3} \sim \mathbb{R}^{6}$. Applying the same procedure outlined in Eqs. (\ref{rv1}) and (\ref{rv2}), we find
\begin{widetext}
\begin{eqnarray}
\varphi _{1,4}^{\prime } &=&\pm \upsilon _{\Phi ^{\prime }}\left( 
\begin{array}{cccccc}
1 & 0 & 1 & 0 & 1 & 0%
\end{array}%
\right) ^{T}\qquad ,\qquad \varphi _{2,5}^{\prime }=\pm \upsilon _{\Phi
^{\prime }}\left( 
\begin{array}{cccccc}
1 & 0 & -1/2 & -\sqrt{3}/2 & -1/2 & \sqrt{3}/2%
\end{array}%
\right)^{T}  \notag \\
\varphi_{3,6}^{\prime } &=&\pm \upsilon _{\Phi ^{\prime }}\left( 
\begin{array}{cccccc}
1 & 0 & -1/2 & \sqrt{3}/2 & -1/2 & -\sqrt{3}/2%
\end{array}%
\right) ^{T}  \label{ev1}
\end{eqnarray}    
\end{widetext}
where the modulus of each vector is given by $\sqrt{3}\upsilon _{\Phi^{\prime }}$. These vectors satisfy a constraint such that $\sum_{i=6}\varphi _{i}^{\prime }=0$, defining a regular Octahedron with $6$ vertices, $12$ edges, and $8$ faces. 
Proving the properties of the Octahedron using the six vacua in Eq. (\ref{ev1}) is a straightforward task. To illustrate this, let us consider the fact that all the edges of an Octahedron possess equal lengths. By examining the right triangular face defined by vertices $\varphi _{1}^{\prime }$, $\varphi_{2}^{\prime }$ and $\varphi _{3}^{\prime }$ as depicted in Fig. \ref{fig2}, we can deduce that $d_{12}^{\prime }=d_{13}^{\prime }=d_{23}^{\prime }=\sqrt{6}\upsilon _{\Phi ^{\prime }}$, where $d_{12}^{\prime }=\norm{\varphi_{2}^{\prime }-\varphi _{1}^{\prime }}$, $d_{13}^{\prime}=\norm{\varphi _{3}^{\prime }-\varphi _{1}^{\prime }}$ and $d_{23}^{\prime }=\norm{\varphi _{3}^{\prime }-\varphi _{2}^{\prime}}$. Furthermore, all the diagonals within an Octahedron share the same length. This fact becomes evident when we calculate the lengths of the three diagonals converging at the center of the Octahedron. Thus, following the notations in Fig. \ref{fig2}, we find that $d_{14}^{\prime} = d_{25}^{\prime }=d_{36}^{\prime }=2\sqrt{3}\upsilon _{\Phi ^{\prime }}$ where $d_{14}^{\prime }=\norm{\varphi _{4}^{\prime }-\varphi_{1}^{\prime }}$, $d_{25}^{\prime }= \norm{\varphi_{5}^{\prime }-\varphi_{2}^{\prime }}$ and $d_{36}^{\prime} = \norm{\varphi _{6}^{\prime }-\varphi_{3}^{\prime}}$.
\subsubsection{DWs for $S_{4}\rightarrow Z_{3}$ breaking}
\begin{figure}[!t]
\centering
\includegraphics[width=0.8\linewidth]{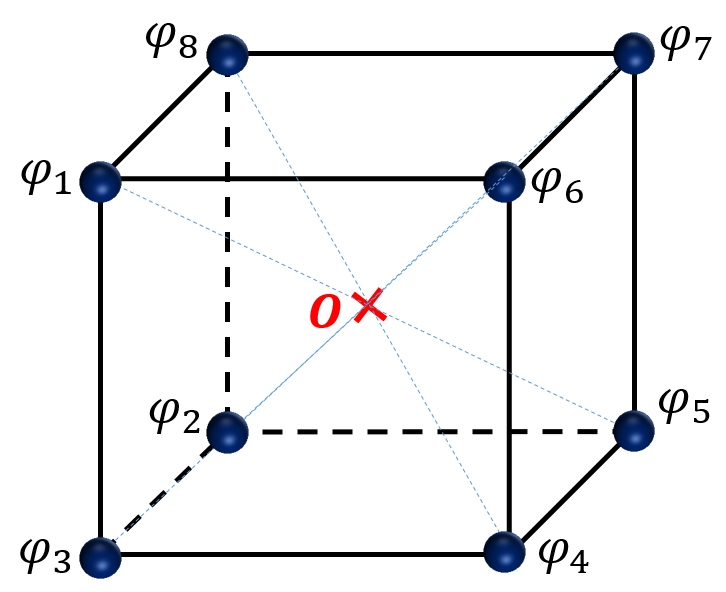}
\caption{A regular cube viewed from the flavon space $\protect\zeta $. The eight vertices represent the eight degenerate vacua, while the edges that connect them stand for domain walls.}
\label{fig1}
\end{figure}

As mentioned in the previous subsection, the breaking pattern $%
S_{4}\rightarrow Z_{3}$ is realized by the flavor triplet $\Phi $ acquiring its VEV along the direction $\left\langle \Phi \right\rangle =\upsilon_{\Phi }(1,0,0)^{T}$ and can be expressed using $G_{f}^{\prime }=Z_{2}^{\mathcal{S}}\times Z_{2}^{\mathcal{U}}\rtimes Z_{2}^{\mathcal{TST}%
^{2}}\rtimes Z_{3}^{\mathcal{T}}$ such that like
\begin{equation}
Z_{2}^{\mathcal{S}}\times Z_{2}^{\mathcal{U}}\rtimes Z_{3}^{\mathcal{T}}\rtimes Z_{2}^{\mathcal{TST}^{2}}\overset{\left\langle \Phi \right\rangle }{\longrightarrow }Z_{3}^{\mathcal{T}},
\end{equation}
where the broken part is given by $\Sigma_{8}=\left( Z_{2}^{\mathcal{S}}\times Z_{2}^{\mathcal{U}}\right) \rtimes Z_{2}^{\mathcal{TST}^{2}}$. The order of $\Sigma_{8}$ is eight which means that the initial $S_{4}$ invariant vacuum gets now split into eight vacua with the same energy $\left\langle \Phi \right\rangle _{i}\equiv \varphi_{i}$ with $i=1,...,8$.
These vacua are situated within the flavon space $\zeta $ and collectively establish the vertices of a regular cube, as depicted in figure \ref{fig1}. The process of determining the coordinates of these eight vacua is straightforward. We know that one of them is just the $Z_{3}^{\mathcal{T}}$-invariant vacuum used for the breaking of $S_{4}$ down to $Z_{3}^{\mathcal{T}%
}$ denoted by $\varphi _{1}=\left\langle \Phi
\right\rangle $. To derive the remaining seven $\varphi _{i}$ vacua, we act upon $\varphi _{1}$ with the generators of $\Sigma_{8}$ by using the representation matrices of the $S_{4}$ generators provided in table (\ref{t2}) as follows
\begin{widetext}
\begin{eqnarray}
\varphi _{1} &=&\left\langle \Phi \right\rangle ~~,~~\varphi _{2}=\mathcal{S}%
\varphi _{1}~~,~~\varphi _{3}=\mathcal{SU}\left( \mathcal{TST}^{2}\right)
\varphi _{1}~~,~~\varphi _{4}=\mathcal{S}\left( \mathcal{TST}^{2}\right)
\varphi _{1}  \notag \\
\varphi _{5} &=&\mathcal{U}\varphi _{1}~~,~~\varphi _{6}=\mathcal{SU}\varphi
_{1}~~,~~\varphi _{7}=\mathcal{TST}^{2}\varphi _{1}~~,~~~\varphi _{8}=%
\mathcal{U}\left( \mathcal{TST}^{2}\right) \varphi _{1}
\end{eqnarray}    
\end{widetext}
This is expressed explicitly in the complex three-dimensional space $\mathbb{C}^{3}$ as follows
\begin{widetext}
\begin{equation}
\varphi_{1,5}=\pm \upsilon _{\Phi }\left( 
\begin{array}{c}
1 \\ 
0 \\ 
0%
\end{array}%
\right) ~~,~~\varphi _{2,6}=\pm \frac{\upsilon _{\Phi }}{3}\left( 
\begin{array}{c}
-1 \\ 
2 \\ 
2%
\end{array}%
\right) ~~,~~\varphi _{3,7}=\pm \frac{\upsilon _{\Phi }}{3}\left( 
\begin{array}{c}
1 \\ 
-2\omega ^{2} \\ 
-2\omega 
\end{array}%
\right) ~~,~~\varphi _{4,8}=\pm \frac{\upsilon _{\Phi }}{3}\left( 
\begin{array}{c}
-1 \\ 
2\omega  \\ 
2\omega ^{2}%
\end{array}%
\right) 
\end{equation}
\end{widetext}
Being associated with $\Sigma _{8}$, these vacua are related to each other by $\left( Z_{2}^{\mathcal{S}}\times Z_{2}^{\mathcal{U}}\right) \rtimes
Z_{2}^{\mathcal{TST}^{2}}$ transformations. Applying the same procedure outlined in Eqs. (\ref{rv1}) and (\ref{rv2}), we find the following expressions for the direction in $\mathbb{R}^6$
\begin{widetext}
\begin{equation}
\begin{array}{ccc}
\varphi _{1,5}=\pm \upsilon _{\Phi }\left( 
\begin{array}{cccccc}
1 & 0 & 0 & 0 & 0 & 0%
\end{array}%
\right)^{T} & , & \varphi_{2,6}=\pm \frac{\upsilon_{\Phi }}{3}\left(
\begin{array}{cccccc}
-1 & 0 & 2 & 0 & 2 & 0%
\end{array}%
\right) ^{T} \\ 
\varphi _{3,7}=\pm \frac{\upsilon _{\Phi }}{3}\left( 
\begin{array}{cccccc}
1 & 0 & 1 & \sqrt{3} & 1 & -\sqrt{3}%
\end{array}%
\right) ^{T} & , & \varphi _{4,8}=\pm \frac{\upsilon _{\Phi }}{3}\left( 
\begin{array}{cccccc}
-1 & 0 & -1 & \sqrt{3} & -1 & -\sqrt{3}%
\end{array}%
\right) ^{T}%
\end{array}
\label{ev2}
\end{equation}
\end{widetext}
where the modulus of each vector is given by $\Re\upsilon _{\Phi }$. These vectors satisfy a constraint such that $\sum_{i=8}\varphi _{i}=0$, defining a cube with $8$ vertices, $12$ edges, and $6$ faces.
The diagonals of this cube converge at a unique point known as the center of symmetry, also referred to as the barycenter (or the core of the cube) of its eight vertices where the complete $S_{4}$ symmetry is manifest. Notably, it is straightforward to prove that connecting these eight vertices creates a cube, as confirmed by verifying that the $12$ edges interconnecting the $8$ vertices in Fig. \ref{fig1} are all of equal length.
This equality can be easily calculated by fixing one vertex at a time. Given that we know that each vertex in a cube is connected to three edges, consider $\varphi _{1}$ in Fig. \ref{fig1}: it is connected via three edges to $\varphi _{3}$, $\varphi _{6}$, and $\varphi _{8}$. As a result, we find that $d_{13}=d_{16}=d_{18}=\sqrt{\frac{4}{3}}\upsilon _{\Phi }$, where $d_{13}=\norm{\varphi _{3}-\varphi_{1}}$, $d_{16}=\norm{\varphi _{6}-\varphi _{1}}$ and $d_{18}=\norm{\varphi_{8}-\varphi_{1}}$.

\begin{figure}[!t]
\centering
\includegraphics[width=0.8\linewidth]{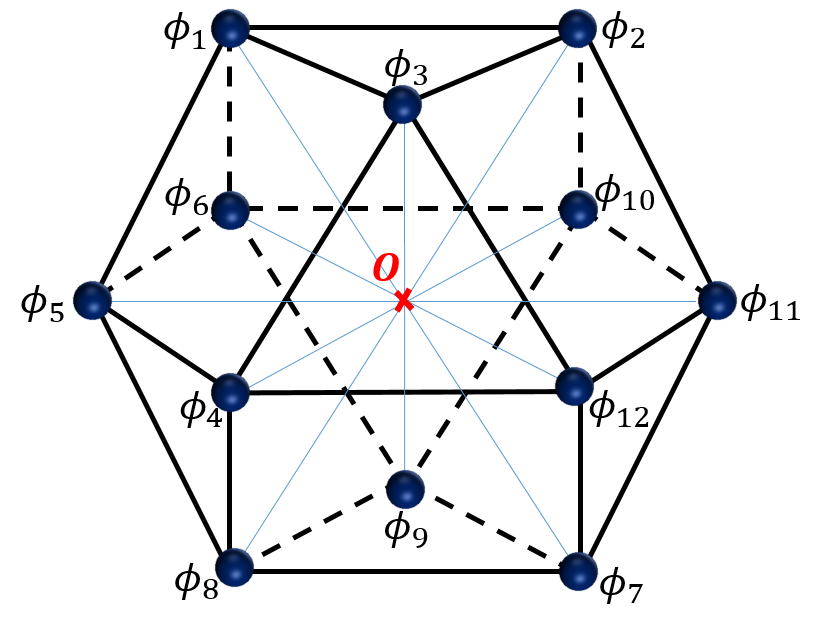}
\caption{A regular Cuboctahedron viewed from the flavon space $\protect\zeta$. The twelve vertices represent the $12$ degenerate vacua, while the edges that connect them stand for domain walls.}
\label{fig3}
\end{figure}

\subsubsection{DWs for $S_{4}\rightarrow Z_{2}$ breaking}
\label{s4z2}
As described previously, the breaking pattern $S_{4}\rightarrow Z_{2}$ can be realized by the flavon triplet\ $\Phi \equiv \mathbf{3}_{(-1,0,-1)}$ acquiring a VEV along the direction $\left\langle \Phi \right\rangle = \upsilon _{\Phi }(2,-1,-1)^{T}$. This breaking can be expressed using $G_{f}^{\prime }$ as follows 
\begin{equation}
Z_{2}^{\mathcal{S}}\times Z_{2}^{\mathcal{U}}\rtimes Z_{3}^{\mathcal{T}}\rtimes Z_{2}^{\mathcal{SU}}\overset{\left\langle \Phi \right\rangle }{%
\longrightarrow }Z_{2}^{\mathcal{SU}},
\end{equation}
where the broken part is given by $Z_{2}^{\mathcal{S}}\times Z_{2}^{\mathcal{U}}\rtimes Z_{3}^{\mathcal{T}}$ which is isomorphic to the alternating $A_{4}$ group of order $12$. Therefore, the initial $S_{4}$ invariant vacuum gets now split into twelve vacua with same energy $\left\langle \Phi \right\rangle _{i}\equiv \phi _{i}$ with $i=1,...,12$. These vacua are located within the flavon space $\zeta $ and collectively establish the vertices of an Archimedean solid known as Cuboctahedron as depicted in figure \ref{fig3}. One of the vertices defining this solid is the $Z_{2}^{\mathcal{SU}}$-invariant vacuum which we denote as $\phi _{1}=\left\langle
\Phi \right\rangle $. As in the previous cases, we act upon $\phi _{1}$ with the generators of $Z_{2}^{\mathcal{S}}\times Z_{2}^{\mathcal{U}}\rtimes Z_{3}^{\mathcal{T}}$ to derive the remaining twelve $\phi _{i}$ vacua
\begin{widetext}
\begin{eqnarray}
\phi _{1} &=&\left\langle \Phi \right\rangle ~~,~~\phi _{2}=\mathcal{T}\phi
_{1}~~,~~\phi _{3}=\mathcal{T}^{2}\phi _{1}~~,~~\phi _{4}=\mathcal{TST}%
^{2}\phi _{1}~~,~~\phi _{5}=\mathcal{ST}^{2}\mathcal{S}\phi _{1}~~,~~\phi
_{6}=\mathcal{STS}\phi _{1}  \notag \\
\phi _{7} &=&\mathcal{U}\phi _{1}~~,~~\phi _{8}=\mathcal{TU}\phi
_{1}~~,~~\phi _{9}=\mathcal{T}^{2}\mathcal{U}\phi _{1}~~,~~\phi _{10}=%
\mathcal{TST}^{2}\mathcal{U}\phi _{1}~~,~~\phi _{11}=\mathcal{ST}^{2}%
\mathcal{SU}\phi _{1}~~,~~\phi _{12}=\mathcal{STSU}\phi _{1}
\end{eqnarray}%
\end{widetext}
In the complex three-dimensional space $\mathbb{C}^{3}$, these transformations lead to
\begin{widetext}
\begin{eqnarray}
\phi _{1,7} &=&\pm \upsilon _{\Phi }\left( 
\begin{array}{c}
2 \\ 
-1 \\ 
-1%
\end{array}%
\right) ~,~\phi _{2,8}=\pm \upsilon _{\Phi }\left( 
\begin{array}{c}
2 \\ 
-\omega ^{2} \\ 
-\omega 
\end{array}%
\right) ~,~\phi _{3,9}=\pm \upsilon _{\Phi }\left( 
\begin{array}{c}
2 \\ 
-\omega  \\ 
-\omega ^{2}%
\end{array}%
\right)   \notag \\
\phi _{4,10} &=&\pm \upsilon _{\Phi }\left( 
\begin{array}{c}
0 \\ 
\omega ^{2}-\omega  \\ 
-\omega ^{2}+\omega 
\end{array}%
\right) ~,~\phi _{5,11}=\pm \upsilon _{\Phi }\left( 
\begin{array}{c}
0 \\ 
\omega ^{2}-1 \\ 
\omega -1%
\end{array}%
\right) ~,~\phi _{6,12}=\pm \upsilon _{\Phi }\left( 
\begin{array}{c}
0 \\ 
\omega -1 \\ 
\omega ^{2}-1%
\end{array}%
\right) 
\label{z2vacua}
\end{eqnarray}%
\end{widetext}
Applying the same procedure outlined in Eqs. (\ref{rv1}) and (\ref{rv2}), we find
\begin{widetext}
\begin{eqnarray}
\phi _{1,7} &=&\pm \upsilon _{\Phi }\left( 
\begin{array}{cccccc}
2 & 0 & -1 & 0 & -1 & 0%
\end{array}%
\right) ^{T}~,\phi _{2,8}=\pm \upsilon _{\Phi }\left( 
\begin{array}{cccccc}
2 & 0 & 1/2 & \sqrt{3}/2 & 1/2 & -\sqrt{3}/2%
\end{array}%
\right) ^{T}  \notag \\
\phi _{3,9} &=&\pm \upsilon _{\Phi }\left( 
\begin{array}{cccccc}
2 & 0 & 1/2 & -\sqrt{3}/2 & 1/2 & \sqrt{3}/2%
\end{array}%
\right) ^{T}~,\phi _{4,10}=\pm \upsilon _{\Phi }\left( 
\begin{array}{cccccc}
0 & 0 & 0 & -\sqrt{3} & 0 & \sqrt{3}%
\end{array}%
\right) ^{T}  \label{ev3} \\
\phi _{5,11} &=&\pm \upsilon _{\Phi }\left( 
\begin{array}{cccccc}
0 & 0 & -3/2 & -\sqrt{3}/2 & -3/2 & \sqrt{3}/2%
\end{array}%
\right) ^{T}~~,\phi _{6,12}=\pm \upsilon _{\Phi }\left( 
\begin{array}{cccccc}
0 & 0 & -3/2 & \sqrt{3}/2 & -3/2 & -\sqrt{3}/2%
\end{array}%
\right) ^{T}  \notag
\end{eqnarray}
\end{widetext}
where the modulus of each vector is given by $\sqrt{6}\upsilon _{\Phi }$. These vectors satisfy a constraint such that $\sum_{i=12}\phi _{i}=0$, defining a regular Cuboctahedron with $12$ vertices, $24$ edges, and $14$ faces (6 square faces and 8 triangular faces).

The properties of the regular Cuboctahedron are easily verified through the real vectors in Eq. (\ref{ev3}). For example, we know that all the edges in this Fig. (\ref{fig3}) are of equal length which we can verify by calculating, for instance, the distances between the edges of the square face $\phi_{1}\phi _{3}\phi_{4}\phi_{5}$ where we find that $d_{13}=d_{34}=d_{45}=d_{51}=\sqrt{6}\upsilon _{\Phi }$ where $d_{ij} = \norm{\phi_i - \phi_j}$.

\begin{table}[!t]
\setlength\tabcolsep{8pt}
\begin{center}
\begin{adjustbox}{max width=0.95\textwidth}
\begin{tabular}{l c c c c c c}
\noalign{\hrule height 1pt}
Fields & $L_{iL}$ & $H$ & $\Delta $ & $\Phi $ & $\sigma $ & $\rho $ \\ 
\noalign{\hrule height 1pt}
$SU(2)_{L}$ & $2$ & $2$ & $3$ & $1$ & $1$ & $1$ \\
$S_{4}$ & $\mathbf{3}$ & $\mathbf{1}$ & $\mathbf{1}^{\prime }$ & $\mathbf{3}$ & $\mathbf{2}$ & $\mathbf{1}^{\prime }$ \\ 
\noalign{\hrule height 1pt}
\end{tabular}%
\end{adjustbox}
\end{center}
\caption{$SU(2)_L$ and $S_4$ transformations of the particle content of the model.}
\label{toy}
\end{table}

\section{A Resolution for the DW problem within a neutrino toy model}
\label{sec:Model}
In this section, we study viable solutions to the DW problem within a toy model that incorporates $S_4$ flavor symmetry. We analyze the neutrino phenomenology in this model, comment on the solutions using high-dimensional operators and at the end we qualitatively discuss the GWs arising from DWs in this case. \\

\subsection{Neutrino phenomenology}
We illustrate the DW problem and its solution using a toy model that is based on the type-II seesaw mechanism with an $S_4$ flavor symmetry. In this study, we will focus solely on the neutrino sector and ignore the charged lepton sector. This is due to the fact that in minimal flavor models, the flavon fields cannot be used simultaneously in the neutrino and charged lepton sectors while preserving the correct vacuum alignment \cite{D9,D10}. Consequently, these two sectors are decoupled from each other, including the scalar potentials of the flavons used in both sectors. Usually, this is achieved by acquiring additional $Z_{N}$ symmetries to control the model Lagrangian and prevent unwanted couplings. \newline
The particle content in this toy model consists of $L_{iL}=(\nu_{iL},\ell_{iL})^T$, $H =(\phi^+,\phi^0)^T$, and $\Delta = (\Delta^{++},\Delta^+,\Delta^0)^T$   
with $L_{iL}, H, $ and $\Delta$ being the lepton doublet, Higgs doublet and scalar triplet. The scalar triplet is responsible for the generation of small neutrino masses through the type-II seesaw mechanism. In addition, we extend the particle content of the SM with three gauge singlet flavon fields ($\Phi, \sigma, \rho$) to achieve the SSB of the $S_4$ symmetry and ensure a Trimaximal mixing for the neutrino mass texture. The transformations of these fields under $SU(2)_L \times S_4$ are summarized in table \ref{toy}.

By using these field transformations, the most general Lagrangian invariant under $SU(2)_L \times U(1)_Y \times S_4$ is given by
\begin{widetext}
\begin{equation}
{\cal L} \equiv \frac{\lambda_\Phi}{\Lambda} \bigg(\bar{L}_{iL}^c i\sigma_2 \Delta L_{jL} \bigg) \Phi + \frac{\lambda_\sigma}{\Lambda} \bigg(\bar{L}_{iL}^c i\sigma_2 \Delta L_{jL} \bigg) \sigma + \frac{\lambda_\rho}{\Lambda} \bigg(\bar{L}_{iL}^c i\sigma_2 \Delta L_{jL} \bigg) \rho + {\rm h.c.},
\end{equation}%
\end{widetext}
where $\Lambda$ is a cutoff scale, $i,j$ are generation indices, and $\lambda_\Phi, \lambda_\sigma$ and $\lambda_\rho$ are independent parameters associated with the flavon fields $\Phi, \sigma$ and $\rho$ respectively. The $S_{4}$ breaking to one of its $Z_{2} $ subgroups is realized when the flavon fields acquire VEVs along the directions
\begin{eqnarray}
\left\langle \Phi \right\rangle &=& \upsilon _{\Phi }(1,1,1)^{T}, \nonumber \\ 
\left\langle \sigma \right\rangle &=& \upsilon _{\sigma }\left( 1,0\right)^{T}, \\
\left\langle \rho \right\rangle &=& \upsilon _{\rho}. \nonumber 
\end{eqnarray}
Using the $S_{4}$ tensor product rules given in the appendix, we obtain the following neutrino mass matrix
\begin{equation}
m_{\nu }=\upsilon _{\Delta }\left( 
\begin{array}{ccc}
2a+b & -a-c & -a \\ 
-a-c & 2a & -a+b \\ 
-a & -a+b & 2a-c%
\end{array}%
\right),
\end{equation}%
where $a, b, c$ are parameters that can be expressed in terms of the VEVs and $\lambda_i$ 
\begin{eqnarray}
    a = \frac{\lambda _{\Phi }\upsilon _{\Phi }}{\Lambda },~b=\frac{\lambda _{\rho }\upsilon _{\rho }}{\Lambda }, ~c=\frac{\lambda_{\sigma}\upsilon _{\sigma }}{\Lambda}.
\end{eqnarray}
In order to account for {\it CP} violation in the neutrino sector, it is necessary for the parameter $c$ to be a complex valued parameter, {\it i.e.} $c \rightarrow |c|e^{i\phi _{c}}$ where $\phi_{c}$ is a {\it CP} violating phase. In the case where $c=0$, $m_\nu$ exhibits the $\mu-\tau$ symmetry \cite{D11,D12,D13,D14,D15} which yields to a Tribimaximal mixing matrix (TBM) \cite{D16}, known for its conservation of $CP$ symmetry. Therefore, the presence of the parameter $c$ breaks this $\mu -\tau$ symmetry giving rise to a neutrino mass matrix characterized by the magic symmetry \cite{D17}. This term alludes to the intriguing property that the sum of elements within any row or column of $m_{\nu}$ remains constant and that $m_{\nu }$ is diagonalized by the well-known Trimaximal mixing matrix ($U_{{\rm TM}_2}$) which is expressed in terms of an arbitrary angle $\theta $ and a phase $\eta$ which will be related to the neutrino oscillation parameters. Since the parameter $c$ is responsible for small deviations from the TBM texture, the modulus of $c$ must satisfy: $|c| < a, b$.
Accordingly, by diagonalizing $m_{\nu }$ we find the following eigenvalues valid up to corrections of order $\mathcal{O}(|c|^{2})$ 
\begin{eqnarray}
    |m_1| &=& v_\Delta \sqrt{(3 a + b)(3 a + b + |c|\cos\phi_c)}, \nonumber \\
    |m_2| &=& v_\Delta \sqrt{b^2 - 2 b |c|\cos\phi_c}, \\
    |m_3| &=& v_\Delta \sqrt{(3a -b)(3 a - b - |c|\cos\phi_c)}. \nonumber
    \label{nm}
\end{eqnarray}
The diagonalization of $m_{\nu }$ by $U_{{\rm TM}_2}$ induces relations between the model parameters $\{a,b,|c|,\phi _{c}\}$ and the Trimaximal mixing parameters $\theta$ and $\eta$ which are found to be
\begin{eqnarray}
\tan 2\theta &=& -\frac{\left\vert c\right\vert \sqrt{9a^{2}\cos ^{2}\phi_{c}+b^{2}\sin ^{2}\phi _{c}}}{\sqrt{3}a\left( b+\left\vert c\right\vert \cos \phi _{c}\right)}, \nonumber \\ \tan \eta &=& \frac{b}{3a}\tan \phi _{c}.
\end{eqnarray}    
From Eq. (\ref{nm}), we derive the expressions for the solar and atmospheric mass-squared differences
\begin{eqnarray}
    \Delta m_{21}^{2}&=&-3\upsilon _{\Delta }^{2}\bigg(|c|\cos \phi _{c}\left(
a+b\right) +a\left( 3a+2b\right)\bigg), \nonumber \\
    \Delta m_{31}^{2}&=&-6\upsilon _{\Delta }^{2}a\bigg(\left\vert c\right\vert \cos
\phi _{c}+2b\bigg).
\end{eqnarray}
The neutrino mixing angles are expressed in the case of ${\rm TM}_2$ with respect to $\theta$ and $\eta$ as follows
\begin{eqnarray}
\sin ^{2}\theta _{12} &=& \frac{1}{3-2\sin ^{2}\theta }, \nonumber \\ \sin^{2}\theta _{13} &=& \frac{2}{3\sin ^{2}\theta}, \nonumber \\ \sin^{2}\theta_{23} &=& \frac{1}{2}-\frac{\sqrt{3}\sin 2\theta}{2\left( 3-\sin ^{2}\theta\right)}\cos\eta.
\end{eqnarray}%
In the context of ${\rm TM}_{2}$, by matching the expression of the Jarlskog parameter from the PDG standard parametrization \cite{D18}, and its expression derived from the Trimaximal mixing matrix $J_{CP}=\left( 1/6\sqrt{3}\right) \sin 2\theta \sin \eta $, we can establish a relationship between $\eta $, the Dirac $CP$ phase and the atmospheric angle
\begin{equation}
\sin 2\theta _{23}\sin \delta _{CP}=\sin \eta.
\end{equation}

There are three different sources that can be used to probe the absolute values of neutrino-mass eigenvalues: \textit{(1)} the sum of the three active neutrino masses from cosmological observations $\sum_{i}m_{i}=m_{1}+m_{2}+m_{3}$, \textit{(2)} direct determination of the neutrino mass by measuring the energy spectrum of the electrons produced in the $\beta $-decay of nuclei which allows to get information on the effective electron antineutrino mass defined by $m_{\beta}=\left( \sum_{i}m_{i}^{2}\left\vert U_{ei}\right\vert ^{2}\right) ^{1/2}$ where $U_{ei}$ are the elements of the first row of the neutrino mixing matrix\footnote{The current limit from Tritium beta decay is given by the KATRIN project, which aims at a detection of $m_{\beta}$ with a sensitivity of $0.2$ ${\rm eV}$
\cite{D19}.}, and \textit{(3)} the search for neutrinoless double beta decay $(0\nu \beta \beta )$ processes whose decay amplitude is proportional to the effective Majorana neutrino mass defined as $\left\vert m_{\beta \beta}\right\vert =\left\vert \sum_{i}U_{ei}^{2}m_{i}\right\vert $. There are many ongoing and upcoming experiments which aim to achieve a sensitivity up
to $0.01$ ${\rm eV}$ for $\left\vert m_{\beta \beta }\right\vert $.
\newline
In the context of the Trimaximal mixing, $U_{ei}$ in $m_{\beta }$ and $\left\vert m_{\beta \beta }\right\vert $ are replaced by the elements of the first row of $U_{{\rm TM}_2}$ while the masses $m_{i}$ are as given in Eq. (\ref{nm}). In this case, two additional phases are introduced and the full mixing matrix can be written
as $U_{{\rm TM}_2} U_{P}$ where $U_{P}={\rm diag}(1,e^{i\alpha _{21}/2},e^{i\alpha
_{31}/2})$ with $\alpha_{21}$ and $\alpha _{31}$ being the two extra Majorana phases. 
\begin{figure*}[!t]
\centering
\includegraphics[width=0.325\linewidth]{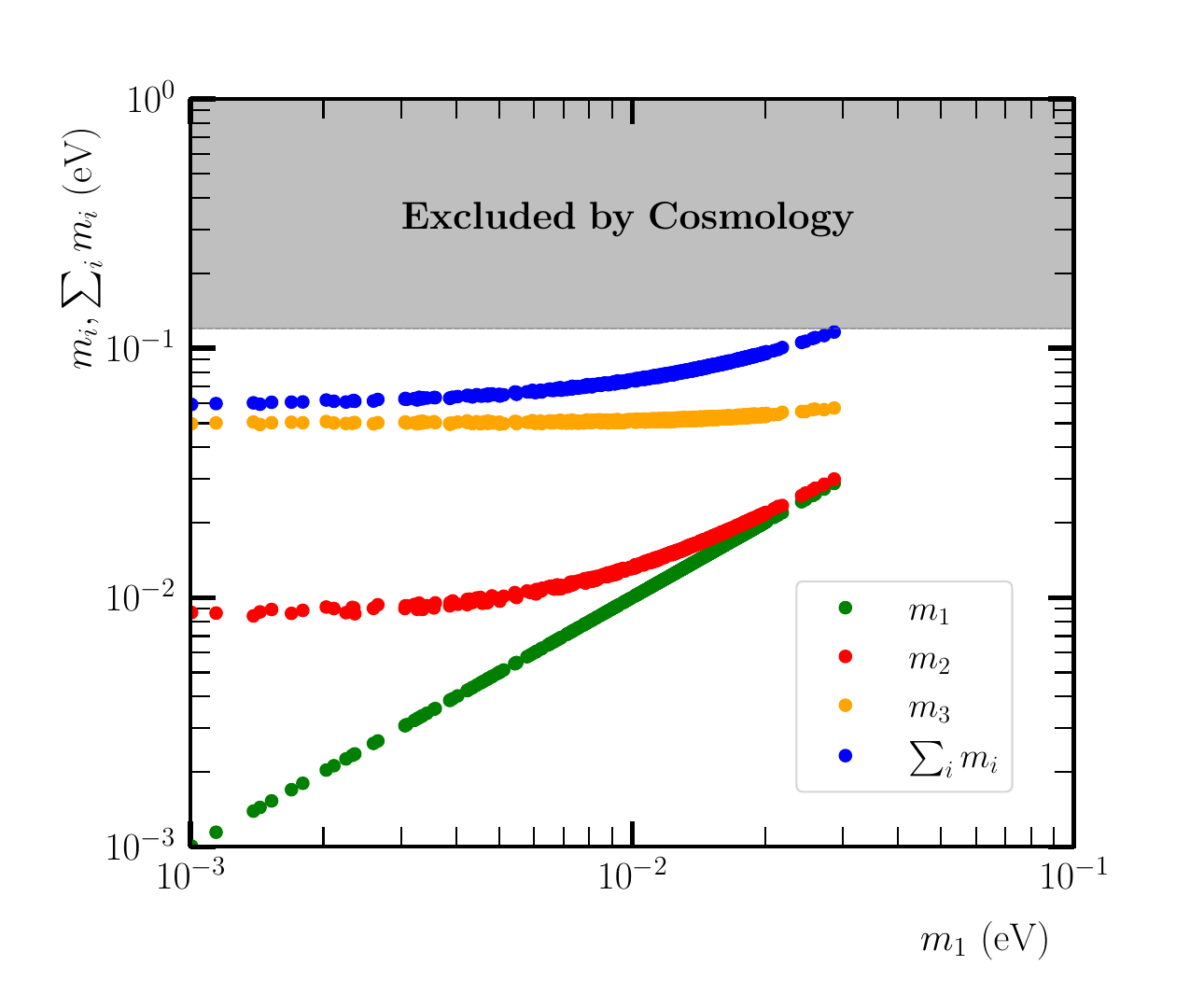}
\includegraphics[width=0.325\linewidth]{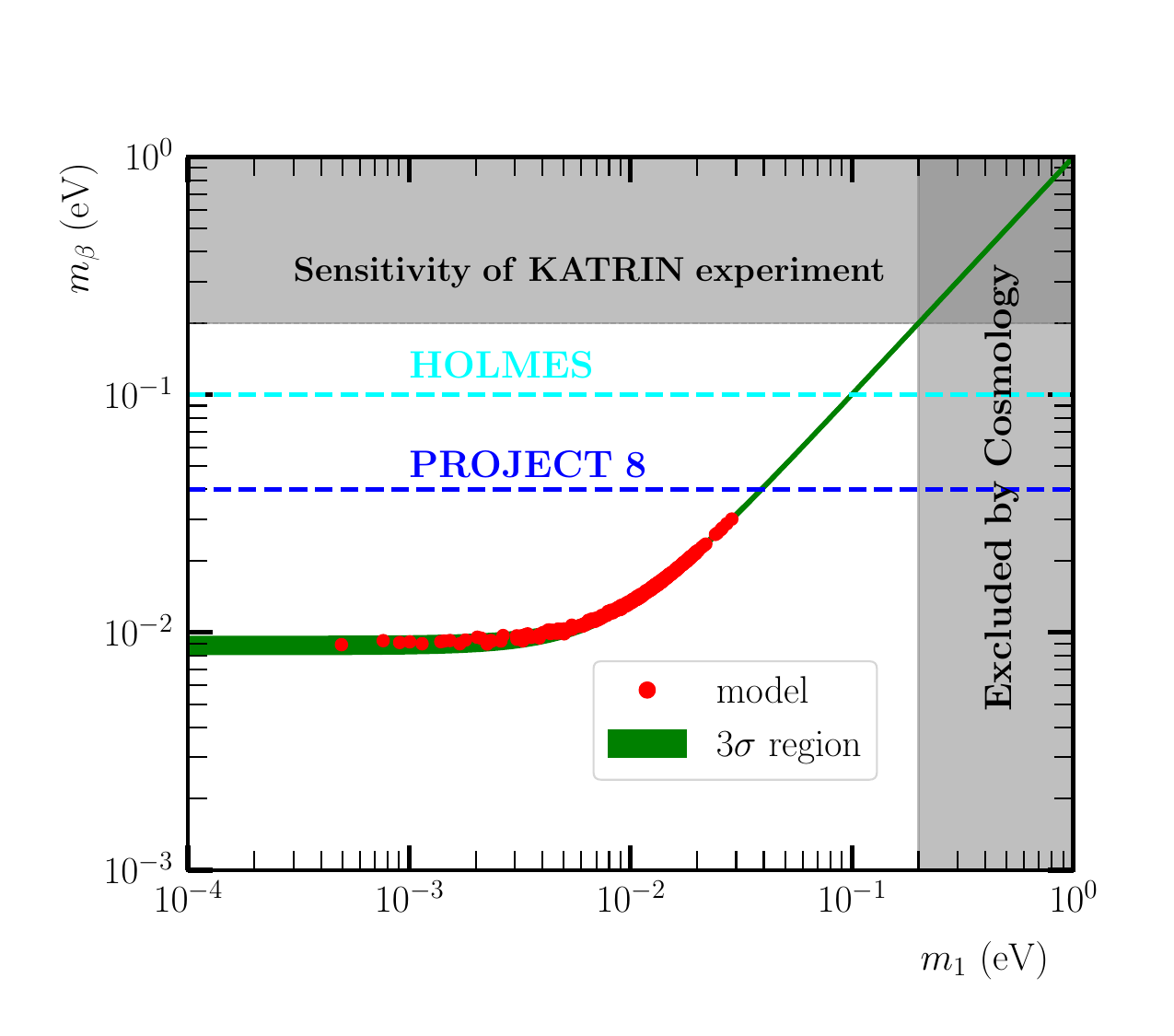}
\includegraphics[width=0.325\linewidth]{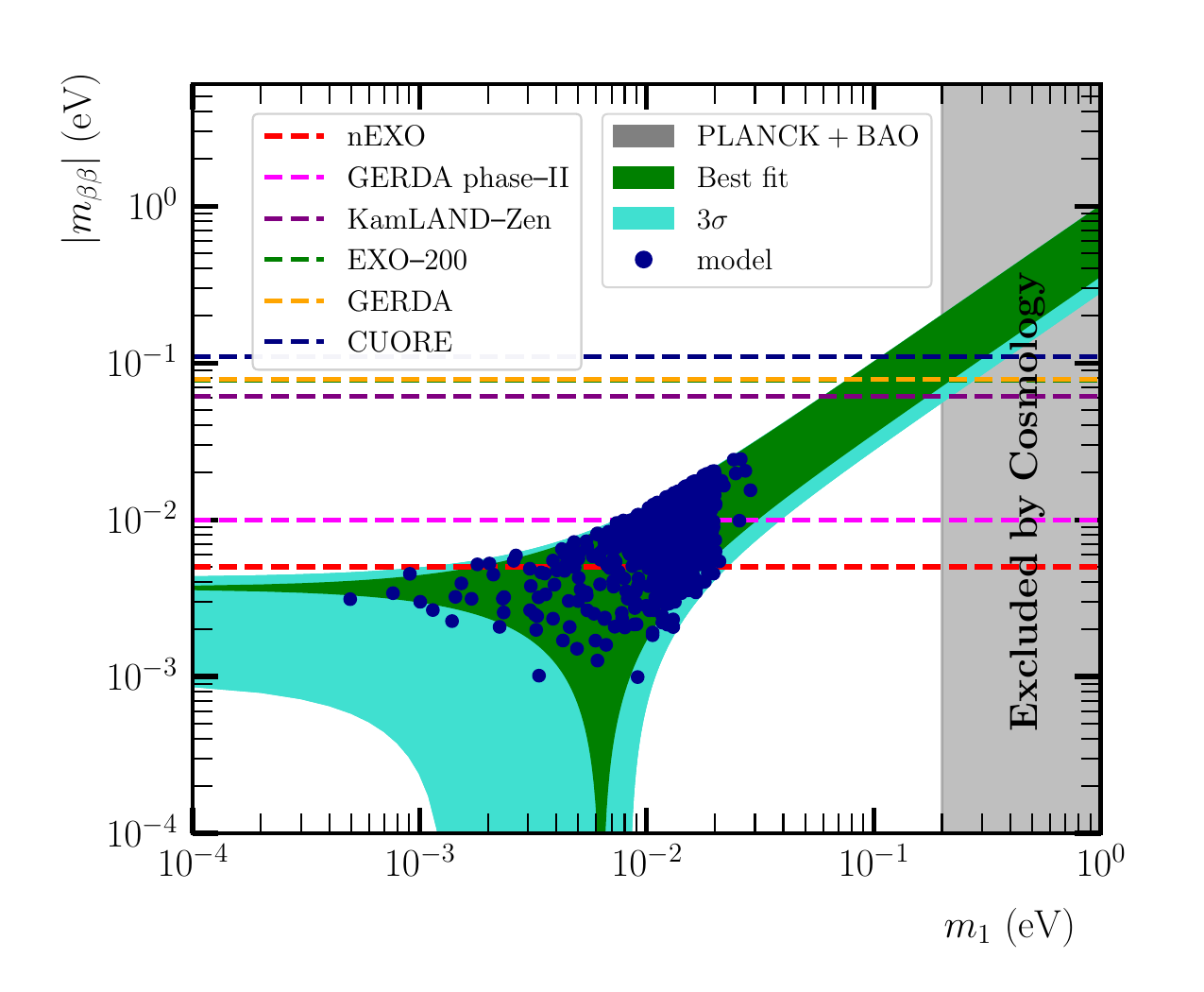}
\caption{Left: prediction for the absolute neutrino masses and their sum $\Sigma m_i$ as a function of $m_1$. Middle: $m_\beta$ as a function of $m_1$ where the vertical and horizontal gray regions are disfavored by Planck and KATRIN collaborations, respectively, while the the horizontal dashed lines represent the future sensitivities on $m_\beta$ from HOLMES and Project 8 collaborations. Right: $|m_{\beta \beta|}$ as a function of $m_1$ where the horizontal dashed lines represent the limits on $|m_{\beta \beta}|$ from current and future $0 \nu \beta \beta$ decay experiments.}
\label{pheno}
\end{figure*}
Since the recent results from NuFiT collaboration show a mild preference for the normal neutrino mass hierarchy \cite{A1}, we will perform a brief numerical analysis of neutrino masses in the case where $m_{1} < m_{2}\lesssim m_{3}$. By taking into account the $3\sigma$ Confidence Level (C.L.) of the oscillation parameters from Ref. \cite{A1}, and the upper bound on the sum of neutrino masses from the latest Planck data $\sum m_{i}<0.12$ eV at 95\% C.L. \cite{D21}, we show in Fig. \ref{pheno} the three neutrino masses and their sum (left panel), $m_{\beta}$ (middle panel) and $\left\vert m_{\beta \beta }\right\vert $ (right panel) as a function of the lightest neutrino mass $m_{1}$. Here, we assume that the angle $\theta$ is allowed to vary in the range $[0,\pi /2]$, and the phases $\eta$, $\phi _{c}$, $\alpha _{21}$ and $\alpha _{31}$ are randomly varied in the range $[0,2\pi ]$ while the parameters $a$, $b$, and $c$ are varied in the range $[-1,1]$. From figure \ref{pheno}, the predicted regions for $m_{i=1,2,3}$, $\sum_{i}m_{i}$, $m_{\beta }$ and $\left\vert m_{\beta \beta}\right\vert $ are as follows%
\begin{widetext}
\begin{eqnarray}
0.00049 &\lesssim& m_{1}[\mathrm{eV}] \lesssim 0.02866, \quad 0.00834 \lesssim m_{2}[\mathrm{eV}] \lesssim 0.02986, \nonumber \\
0.04941 &\lesssim& m_{3}[\mathrm{eV}] \lesssim 0.05754, \quad 0.05884 \lesssim \sum m_{i}[\mathrm{eV}] \lesssim 0.11607, \\ 
0.00889 &\lesssim& 
m_{\beta }[\mathrm{eV}] \lesssim 0.03001, \quad 0.00099 \lesssim \left\vert m_{\beta \beta }\right\vert [\mathrm{eV}] \lesssim 0.02438. \nonumber
\label{pred}
\end{eqnarray}    
\end{widetext}
The predicted upper bound value of $\sum m_{i}$ is close to the result of the Planck collaboration \cite{D21}, while the lower bound ($\sim 0.06$ $\mathrm{eV}$) requires further studies in future cosmological data and may be tested in the future by experiments such as CORE+BAO aiming to reach a $0.062$ eV sensitivity \cite{D22}. For $m_{\beta }$, the obtained values are far from the forthcoming $\beta $-decay experiment sensitivities \cite{D23,D24,D25,D26}, and thus require experiments with improved sensitivities around $0.02$ $\mathrm{eV}$. For $\left\vert m_{\beta \beta }\right\vert $, the horizontal dashed lines in Fig. \ref{pheno} show that the model values are below the current sensitivities for some of the
ongoing $0\nu \beta \beta $\ decay experiments while the anticipated sensitivities of the next-generation experiments such as GERDA Phase II \cite{D27} and nEXO \cite{D28} will cover our model predictions on $\left\vert m_{\beta \beta }\right\vert $. Moreover, it is clear from the mass expressions in Eq. (\ref{nm}) that the tiny neutrino masses constraint the parameter $\upsilon _{\Delta }$, which has been determined to fall within
the range of $0.01803\lesssim \upsilon _{\Delta }\left[ \mathrm{eV}\right] \lesssim 0.04997$.
\subsection{Solution to the DW problem}
\subsubsection{Exploring the DW problem and potential remedies}
As outlined above, DWs are sheet-like topological defects that emerge as a result of the SSB of discrete groups. These two-dimensional structures remain stable when the vacuum manifold, defined by the scalar potential of the theory, is topologically nontrivial. This non-triviality manifests as isolated points corresponding to distinct degenerate ground states within the theory. In the classic example of SSB of the discrete $Z_{2}$ symmetry driven by a real scalar field transforming as $\varphi \rightarrow -\varphi $, the potential has two degenerate vacua at $\varphi=\pm \upsilon $. Thus, the associated manifold is restricted to only two
points. In our toy model where $S_{4}$ breaks down to $Z_{2}$, the manifold is made of twelve disconnected vacua given by the quotient group $S_{4}/Z_{2}$ which corresponds to the broken part $V_{4}\rtimes Z_{3}$ as discussed in section \ref{sec2}. Once the scalar field settle into one of these possible ground states representing a point within the vacuum manifold, depicted graphically as the Cuboctahedron in figure \ref{fig3}, it becomes
impossible to transition to another point within the manifold. The
transitional regions between these degenerate ground states are what we refer to as DWs.

In order to understand why stable DWs are problematic, it is essential to comprehend the cosmological implications of the SSB of a discrete group. According to the standard cosmological model, the initial state of the universe is a state of local thermal equilibrium at extremely high temperature. Typically, spontaneously broken symmetries are restored at this high temperature. However, as the universe goes through cosmic expansion and gradually cools over time, a crucial phase transition takes place when the temperature falls below a critical threshold leading to the SSB of the discrete symmetry. The issue arises from the fact that distinct regions with a
characteristic size of approximately the Hubble radius $H^{-1}$ become causally disconnected. After the symmetry is broken, these regions will be situated in different vacua, separated by DWs. A comprehensive study of the cosmological evolution of a universe containing these DWs was first performed by Zel'dovich, Kobzarev and Okun \cite{A4}, where they concluded that the contribution of stable DWs to the energy density
of the universe would rapidly surpass the contribution from radiation, causing a rapid expansion of the universe that would leave less time for the formation of galaxies and subsequently impact the production rates during Big Bang Nucleosynthesis. Indeed, the energy density of DWs can be expressed as
\begin{equation}
\rho_{\rm DW}= \sigma_{\rm DW}/R,
\end{equation}
where $R$ represents the typical size of the domain walls DWs, and $\sigma_{\rm DW}$ denotes the surface energy density of the walls.
The parameter $R$ is directly proportional to the scale factor $a$, and thus the energy density of domain walls scales as $a^{-1}$, which is significantly lower than the scaling of radiation $a^{-4}$ or matter $a^{-3}$. This means that as time progresses, the energy density of DWs decreases at a slower rate than that of matter or radiation. Eventually, regardless of their initial abundance, their presence at late times would make them the dominant contributors to the total energy density of the universe. This has profound cosmological implications that conflicts observational data, notably leaving significant imprints in the CMB radiation. To elaborate this point, recall that the CMB exhibits an incredibly precise black body radiation \cite{B2}, and any deviation from this spectrum, referred to as CMB spectral distortions, holds invaluable information about the physics of the early universe. Specifically, the primordial fluctuations of matter density, that will later form large scale structures we observe today, leave imprints in the form of temperature
anisotropies in the CMB. Nevertheless, these temperature fluctuations are very small, with the CMB spectrum maintaining remarkable uniformity up to about 1 part in $10^5$, usually expressed as 
\begin{equation}
\frac{\delta T}{T} \lesssim 10^{-5},
\end{equation}
where $\delta T$ denotes the temperature deviation in a given region of the sky, and $T$ denotes the mean temperature of the CMB. With these cosmic evolution facts in mind, a notable challenge that arises with the existence of stable DWs within the observable universe is that they would introduce additional distortions, or excessive anisotropies, in contrast to the CMB
anisotropies that we observe today. This issue emerges particularly in cases where the symmetry breaking scale is around $\upsilon_{\varphi }\gtrsim 1MeV$ as shown in Ref. \cite{A4}. This bound is estimated by equating the density fluctuation $\delta \rho_{\rm DW} /\rho_{\rm DW} $ of a DW to the
temperature fluctuation in the CMB, $\delta T/T$, which are approximately equal at the surface of the last scattering (post-recombination) when the DW spans across the present observable horizon \cite{B3}. The density fluctuation for a DW located between two degenerate vacua is estimated to be 
\begin{equation}
\frac{\delta \rho_{\rm DW}}{\rho_{\rm DW}} \sim 10^{60} \times \bigg(\frac{\upsilon_{\varphi}}{M_{\rm Planck}}\bigg)^3,
\end{equation}
where $M_{\rm Planck} = 1.2 \times 10^{19}$ GeV is the Planck mass. Consequently, the constraint $\delta \rho_{\rm DW}
/\rho_{\rm DW} \sim \delta T/T\lesssim 10^{-5}$ establishes a lower limit on $\upsilon_\varphi$: $\upsilon_{\varphi }\gtrsim$ 1 MeV under which stable DWs pose significant cosmological challenges. Conversely, these DWs do not present any issues if the gravitational redshift of the walls remain well below the detectable threshold for anisotropy constraints, $\delta T/T\lesssim 10^{-5}$. This implies that the energy scale of the phase transition that generates these DWs must be limited to values smaller than $1$ MeV.

Considering the strict constraint imposed on the scale of the SSB mentioned above, different approaches have been proposed in the literature to address the DW issue. These approaches revolve around the concept of unstable DWs that experience early decay, implying their existence
for only a brief period of time \cite{A6}. The most famous approach is inflation \cite{D29}, which is a valid solution provided that the DWs formed prior to the end of inflation, causing them to expand beyond our current observable horizon. In simpler terms, this implies that the discrete symmetry must have already broken before the onset of inflation (the phase transition must occur before the inflation). Consequently, inflation will cause these DWs to expand and disappear from our observable universe, eliminating the associated problem. A second approach for addressing the DW problem assumes that the discrete symmetry connecting the vacuum states as approximate symmetry. This suggests the existence of an explicit symmetry breaking term, which would result in the eventual collapse of the DWs during early cosmic times \cite{B3}. It was shown in Refs. \cite{A4,A5} that this can be achieved by introducing a biased potential that can effectively lift the degeneracy among the vacua. This allows one of the vacuum states to attain a slightly lower energy density than the others, ultimately establishing its dominance in the universe as the true vacuum. Typically, this symmetry breaking term is introduced by hand, although in principle, the soundness of this solution depends on whether explicit breaking naturally arises from some underlying
physics \cite{D32}. Indeed, it was shown in Refs. \cite{D33,D34} that this explicit breaking term can be achieved by taking into account Planck scale gravitational effects. This is realized via the introduction of higher-dimensional operators, denoted as 
\begin{equation}
\frac{1}{M_{\rm Planck}^{n}}O_{n+4} + {\rm h.c}.
\end{equation}
which are suppressed by powers of the Planck mass $M_{\rm Planck}$ in the scalar potential leading to a preference for one of the vacua over the others. This same concept is applicable to our $S_{4}$ toy model, which features twelve degenerate vacua. By introducing a slight bias in favor of one vacuum state over the others, we effectively resolve the DW problem.
\subsubsection{Solving the DW problem within our toy model}
In the toy model outlined above, the $S_{4}$ flavor symmetry is broken into one of the $Z_{2}$ subgroups of $S_{4}$ with the broken part denoted as $K_{4}\rtimes Z_{3}\subset S_{4}$. This breaking is realized by one of the flavons listed in table \ref{toy}. As a result, DWs form, and expand between the boundaries of twelve degenerate vacua $\phi _{i=1,...,12}$ which are distinguished by the transformation of the broken part $K_{4}\rtimes Z_{3}$ as detailed in subsection \ref{s4z2}. These DWs pose a challenge when the scale of the symmetry breaking associated with their formation falls below the inflationary scale which is approximately $10^{16}$ GeV \cite{D35,D36,D37}. In our toy model, there is no inherent constraint preventing the VEVs of the flavons \{$\Phi ,\sigma ,\rho $\} to fall below the inflationary scale threshold. Let us verify this by getting an estimate on the $S_{4}$ breaking scale using, as an example, the expression of $\left\vert m_{2}\right\vert $ given in Eq. (\ref{nm}). The upper bound of
this mass, $\left\vert m_{2}\right\vert \approx 0.0298$ eV, is given in Eq. (\ref{pred}) and is obtained for 
\begin{eqnarray}
\upsilon _{\Delta } &\approx& 0.047~{\rm eV},\quad b\approx 0.827, \nonumber \\ 
\left\vert c\right\vert &\approx& 0.198,\quad \cos\phi_{c}=0.875.    
\end{eqnarray}
From the expressions of $\left\vert c\right\vert =\frac{\lambda _{\sigma}\upsilon _{\sigma }}{\Lambda }$ and $b=\frac{\lambda _{\rho }\upsilon_{\rho }}{\Lambda }$, and assuming for simplicity that the Yukawa couplings are of order $\mathcal{O}(1)$, it is clear that when considering the VEV of flavons $\rho $ and $\sigma $ around the TeV scale, $\Lambda $ will always remain below the inflationary scale. Therefore, we assert that the DWs
created during $S_{4}$ symmetry breaking are inconsistent with the standard cosmology \cite{D38}, and must be prevented. \newline
For this purpose, we adopt the second approach mentioned above and we break explicitly the $K_{4}\rtimes Z_{3}$ subgroup of $S_{4}$ at a high energy scale by using one of the possible Planck-suppressed operators within our toy model. The leading higher dimensional terms induced by gravity and respecting gauge
invariance are of dimension five, thus they are suppressed by one power of $M_{\rm Planck}$. It is worth noting that due to the substantial suppression of these operators, the model remains unchanged at lower energy scales.\newline
There exist several 5-dimensional operators capable of explicitly breaking the $S_{4}$ flavor group. However, it is not necessary to enumerate or employ all of them. As stated in Ref. \cite{D38}, even the tiny higher dimensional symmetry-breaking terms, constrained by powers of the Planck mass, may suffice to address the DW problem. Given our primary focus on $S_{4}$ flavon triplets, we shall employ the flavon triplet $\Phi
\sim \mathbf{3}$. During the SSB of the $S_{4}$ group, this triplet creates DWs that are located on the boundaries of the twelve degenerate vacua as described in Eq. \ref{z2vacua} and depicted in Fig. \ref{fig3}. Therefore, the leading contribution of $\Phi$ in the effective scalar potential may be expressed as
\begin{widetext}
\begin{equation}
\Delta V_{\rm eff}=\frac{\varepsilon _{5}}{M_{\rm Planck}}\left[ \left. \left[ \left(\Phi ^{\dagger }\Phi \right) _{\mathbf{3}}\left( \Phi ^{\dagger }\Phi \right) _{\mathbf{3}}\right] _{\mathbf{3}^{\prime }}\left( \Phi ^{\dagger
}\right) _{\mathbf{3}}\right\vert _{\mathbf{1}^{\prime }}+{\rm h.c.}\right].
\label{sdw}    
\end{equation}
\end{widetext}

Since this term transforms as a nontrivial $S_{4}$ singlet it breaks explicitly the $S_{4}$ group. When the flavon triplet obtains its VEV, $\left\langle \Phi \right\rangle =\left( \upsilon _{\Phi },\upsilon _{\Phi},\upsilon _{\Phi }\right) ^{T}$, the resulting contribution from $\Delta V_{\rm eff}$ becomes sufficient to lift the degeneracy among the twelve degenerate vacua $\phi _{i=1,...,12}$. In such a scenario, the energy-density difference between these vacua can be approximated as 
\begin{eqnarray}
V_{bias} \approx \frac{\varepsilon_5 \upsilon_\Phi^{5}}{M_{\rm Planck}},
\label{eq:DeltaV}
\end{eqnarray}
where $\upsilon _{\Phi}$ represents the scale of spontaneous symmetry breaking of the $S_{4}$ group. Therefore, to estimate the amount of symmetry breaking, it is necessary to determine the values of the bias coefficient $\varepsilon_{5}$. Assuming that this coefficient is real, the general condition for the elimination of DWs, taking into account the operator mentioned in Eq. (\ref{sdw}) can be expressed as follows \cite{D38}
\begin{equation}
\varepsilon_{5}>\left( \frac{\upsilon_{\Phi }}{M_{\rm Planck}}\right) 
\end{equation}
where $\varepsilon_{5}$ is a positive dimensionless parameter that characterizes the effective bias. As previously discussed, there is no inherent constraint preventing SSB of $S_{4}$ from occurring at scales lower than that of inflation. In fact, as the scale of symmetry breaking decreases, the possibility of creating an energy gap among the degenerate vacua becomes increasingly plausible \cite{D39}. For instance, if we consider $\upsilon_{\Phi }$ to be around the TeV scale, say $\upsilon _{\Phi }\sim \mathcal{O}$(TeV), it is sufficient to require $\varepsilon _{5}>10^{-17}$ in order favor one of the vacua over the others.
\subsubsection{Gravitational waves from DWs}
The dimension-five operator in Eq. \ref{sdw} induces a slight bias among degenerate minima, ultimately favoring one true vacuum state. Over time, this bias becomes dynamically significant, accelerating each wall towards its adjacent higher energy vacuum and driving the evolution of the DWs towards their eventual annihilation. The dynamics of these walls can be fully understood if two underlying forces are determined. The first one is the surface tension ($p_{T}$) caused by the curvature and is proportional to the energy per unit area: $p_{T} \sim \sigma_{\rm DW}/R$. This force acts as a surface pressure that straightens the curved walls up to the horizon scale. The second force is the volume pressure which shrinks the false vacuum domains leading to the collapse of DWs and is approximately scaling as $p_{V}\sim v T^{4}$ \cite{L1}, where $v$ is the velocity of the
DWs. This force is equal to the differences in energy density of the vacua and thus may be expressed also as $p_{V}\sim V_{\rm bias} =\left( \varepsilon_{5}\upsilon_{\Phi }^{5}\right) /M_{\rm Planck}$. These two forces compete, and the dynamics of the walls is profoundly influenced by the magnitude of the bias \cite{A6}. DWs annihilate when the pressure $p_{V}$ becomes comparable to the tension force $p_{T}$. This annihilation process is remarkably
energetic, resulting in the emission of stochastic GWs which retain various information on the physics of the early Universe \cite{D39,D40,D41,D42,D43,D44}. The time of annihilation $t_{\rm ann}$ and the corresponding temperature $T_{\rm ann}$ are given by 
\begin{eqnarray}
t_{\rm ann} &\simeq& \frac{\sigma_{\rm DW}}{V_{\rm bias}}, \\
T_{\rm ann} &=& 3 \times 10^{7} \left(\frac{10}{g_{\ast}(T_{\rm ann})}\right)^{1/4}
\left(\frac{V_{\rm bias} \sigma_{\rm DW}^{-1}}{1~{\rm TeV}}\right) ^{1/2}~{\rm TeV}, \nonumber
\end{eqnarray}    
where $g_{\ast }(T_{\rm ann})$ is the effective number of degrees of freedom at $t_{\rm ann}$. In most realistic cases, $(10/g_{\ast})^{1/4}$ is of order one. Using the expression of $V_{\rm bias}$ in Eq. \ref{eq:DeltaV}, we have 
\begin{eqnarray}
t_{\rm ann} &\simeq& \frac{\sigma_{\rm DW} M_{\rm Planck}}{\varepsilon_{5} \upsilon _{\Phi }^{5}}, \\ 
T_{\rm ann} &\simeq& 3\times 10^{7} \left(\frac{\varepsilon_{5}\upsilon_{\Phi}^{5} \sigma_{\rm DW}^{-1} M_{\rm Planck}^{-1}}{1~{\rm TeV}}\right)^{1/2}~{\rm TeV}.  \nonumber  
\end{eqnarray}    
Once the tension force becomes dominant, DWs will enter the scaling regime in which the typical length scales are given by the Hubble radius $H^{-1}$ as mentioned previously. This is the so-called scaling solution, for which the energy density of the DWs evolves as $\rho_{\rm DW} = {\cal A}\sigma _{DW}/t$ where ${\cal A}$ is a dimensionless numerical factor called the area parameter which takes an almost constant value \cite{D44}\footnote{The scaling properties of DWs have been verified through several numerical \cite{D44} and analytical \cite{D48} techniques. For instance, using numerical simulations in a model with two degenerate vacua ($Z_{2}$-invariant scalar potential), it has been found that the parameter $\mathcal{A}$ approximates to $\mathcal{A}\simeq 0.8\pm 0.1$. In models with $N>2$ degenerate vacua, this parameter is usually estimated based on the $Z_{2}$ example and its value increases proportionally with $N$ \cite{D49,D50}. In Ref. \cite{L2}, the value of $\mathcal{A}$ was estimated as $\mathcal{A}\simeq 0.8\times (3/2)$ in a model with three degenerate vacua (i.e. $Z_{3}$ invariant NMSSM).}. In this scaling regime, the peak amplitude of the stochastic GWs spectrum at the present time $t_{0}$ can be expressed as \cite{L2,D44}
\begin{eqnarray}
\Omega _{\rm GW} h^2 (t_{0})|_{\rm peak} &\simeq& 5.2\times 10^{-20}
\tilde{\epsilon}_{\rm gw} {\cal A}^4 \left(\frac{10.75}{g_{\ast }(T_{\rm ann})}
\right)^{1/3} \nonumber \\ 
&\times& \left( \frac{\sigma_{\rm DW}}{1~{\rm TeV}^3}\right)^{4}\left( \frac{%
1~{\rm MeV}^{4}}{V_{\rm bias}}\right)^{2}, 
\label{gwa}
\end{eqnarray} 
where $\tilde{\epsilon}_{\rm gw}$ is an efficiency parameter determined in numerical simulations to be equal to $\tilde{\epsilon}_{\rm gw}\simeq 0.7$ \cite{D44}. Beyond the peak, the amplitude of GWs spectrum induced by collapsing
DWs is given by
\begin{equation}
\Omega _{\rm GW}h^{2}\simeq \left. \Omega _{\rm GW} h^{2}\right\vert _{\rm peak}\times
\left\{ 
\begin{array}{c}
\left( \frac{f}{f_{\rm peak}}\right) ^{3},\quad f<f_{\rm peak} \\ 
\frac{f}{f_{\rm peak}},\quad f>f_{\rm peak}%
\end{array}%
\right. 
\end{equation}
where the peak frequency is found to be 
\begin{eqnarray}
f\left( t_{0}\right)_{\rm peak} &\simeq& 3.99\times 10^{-19}\times {\cal A}%
^{-1/2} \times \left( \frac{1~{\rm TeV}^{3}}{\sigma _{\rm DW}}\right)^{1/2} \nonumber \\ 
&\times& \left( \frac{V_{\rm bias}}{1~{\rm MeV}^{4}}\right)^{1/2}~{\rm Hz}.  
\label{freq}    
\end{eqnarray}
It is clear from Eqs. (\ref{gwa}) and (\ref{freq}) that both the amplitude and the peak frequency of the GWs depend on two parameters: $\sigma_{\rm DW}$ and $V_{\rm bias}$. Since both $\sigma_{\rm DW}$ and $V_{\rm bias}$ depend on the scale of the SSB of $S_4$ symmetry ($\upsilon_\Phi$), the spectrum of $\Omega_{\rm GW} h^2$ can be used to constrain the SSB of $S_4$ within this model. It is worth noting that the same scale ($\upsilon_\Phi$) controls some of the neutrino predictions. Therefore detailed analyses will shed some lights on the connections between flavor symmetries, neutrino mass and the DW problem through GWs.

Our primary goal in this paper is to discuss the theoretical and geometric aspects of the DW problem arising from the spontaneous breaking of the $S_{4} $ flavor symmetry into its potential residual symmetries. While we have outlined the subtleties of GWs using parameters from our toy model, more involved analyses of GW probes of DWs within models based on non-Abelian discrete groups necessitate further investigations of the wall network that relies in part on dedicated Monte Carlo simulations which we leave for a future work.
\section{Summary and Conclusions}
\label{concl}
In this work, we have studied the formation of cosmic DWs from the SSB of the $S_{4}$ discrete group. This breaking occurs when gauge singlet flavon fields acquire VEVs giving rise to multiple distinct degenerate vacua separated by energy barriers which establish a network of DWs. These non-Abelian discrete groups are
widely used in flavor model building, primarily because of their ability to give predictions that are in good agreement with neutrino oscillation data. In these models, a crucial point is that there are distinct preserved residual symmetries in the neutrino and charged lepton sectors after the SSB. This distinction arises from the fact that the flavon configurations that predict specific fermion mass structures exhibit different VEV directions in both sectors. Consequently, the number of vacua is contingent upon both the order of the residual symmetry and the order of the underlying flavor group. In the case of the $S_{4}$ flavor symmetry, we have investigated three possible breaking patterns that are phenomenologically viable: $S_{4}\rightarrow K_{4}$, $S_{4}\rightarrow Z_{3}$, and $S_{4}\rightarrow Z_{2}$. The associated broken subgroups are given by the non-Abelian groups $S_{3}\simeq S_{4}/K_{4},$ $\Sigma (8)\simeq S_{4}/Z_{3}$, and $A_{4}\simeq S_{4}/Z_{2}$ respectively. Therefore, the number of vacua for each breaking pattern is given by the order of $S_{3}$, $\Sigma (8)$ and $A_{4}$, namely 6, 8 and 12 respectively. The challenges associated with presenting DWs in cases of multiple vacua have been addressed through a novel approach where we have depicted the $S_{4}$ DW networks by representing each breaking pattern's multiple vacua as vertices. Each vertex is expressed through vector coordinates in flavon space. The resulting outcome consists of Archimidean or Platonic solids, created by connecting vertices that represent degenerate vacua with edges that symbolize DWs. This provides a clear visualization of the intricate network structure.

On the other hand, we addressed the challenge of stable DWs by introducing an $S_{4}$ toy model wherein neutrino masses are generated through the type-II seesaw mechanism. After the SSB of gauge and flavor symmetries, the resulting neutrino mass matrix exhibits a magic symmetry, and it is diagonalized by the Trimaximal mixing matrix known to align with neutrino oscillation data. Moreover, we find that the obtained predictions concerning the sum of the three active neutrino masses $\sum m_{i}$, and the
effective Majorana mass $m_{\beta \beta }$ can be probed by future
experiments. The breaking pattern in this toy model is given by $%
S_{4}\rightarrow Z_{2}$, which gives rise to twelve degenerate vacua, and to resolve the DW problem, we have adopted the approach of explicit symmetry breaking through the introduction of a Planck-suppressed operator induced by gravity into the scalar potential. This explicit breaking induces
a bias among the twelve degenerate vacua, parameterized as $\varepsilon_{5}\upsilon _{\Phi }^{5}$, where $\upsilon _{\Phi }$\ is the scale of the SSB of the $S_{4}$ group. Lifting the degeneracy of the multiple minima of the scalar potential while favoring one true vacuum state is obtained by putting constraints on the dimensionless coefficient $\varepsilon _{5}$ that characterized the bias where we find that for $\upsilon _{\Phi }\sim \mathcal{O}$(1) TeV, it is sufficient to require $\varepsilon _{5}>10^{-17}$ in order to drive the evolution of the walls towards their annihilation.\newline
An interesting outcome of the annihilation of DWs is the production of GWs, offering a potential avenue for probing these models through ongoing and planned GW experiments. However, accurately describing gravitational wave emission in models based on non-Abelian discrete groups necessitates a thorough examination of the parameters influencing the evolution of the formed walls. This involves undertaking dedicated numerical simulations, presenting a complex scenario in contrast to the well-established examples found in literature with $Z_{2}$ symmetry. A thorough examination of this topic is reserved for future investigations

\acknowledgments
The work of A.J. is supported by the Institute for Basic Science (IBS) under the project code, IBS-R018-D1. The work of M.A.L and S.N. is supported by the United Arab Emirates University (UAEU) under UPAR Grant No. 12S093.

\appendix

\begin{table}[!tbp]
\setlength\tabcolsep{12pt}
\begin{center}
\begin{adjustbox}{max width=0.95\textwidth}
\begin{tabular}{l c c c c c}
\noalign{\hrule height 1pt}
Classes & $\chi _{\mathbf{1}}$ & $\chi _{\mathbf{1}^{\prime }}$ & $\chi _{%
\mathbf{2}}$ & $\chi _{\mathbf{3}}$ & $\chi _{\mathbf{3}^{\prime }}$ \\ 
\noalign{\hrule height 1pt}
$\ \ \ \mathcal{C}_{1}$ & $1$ & $1$ & $2$ & $3$ & $3$ \\ 
$\ \ \ \mathcal{C}_{3}$ & $1$ & $1$ & $2$ & $-1$ & $-1$ \\ 
$\ \ \ \mathcal{C}_{6}$ & $1$ & $-1$ & $0$ & $-1$ & $1$ \\ 
$\ \ \ \mathcal{C}_{8}$ & $1$ & $1$ & $-1$ & $0$ & $0$ \\
$\ \ \ \mathcal{C}_{6}^{\prime }$ & $1$ & $-1$ & $0$ & $1$ & $-1$ \\ 
\noalign{\hrule height 1pt}
\end{tabular}
\end{adjustbox}
\end{center}
\caption{Characters of the irreducible representations of the $S_4$ group.}
\label{tab:characters}
\end{table}%

\section{The $S_{4}$ group}
The flavor group $S_{4}$ is the permutation group of four distinct objects. It contains $\mathbf{24}$ elements which can be generated by three generators $\mathcal{T}$, $\mathcal{S}$ and $\mathcal{U}$ satisfying $\mathcal{S}^{2}=\mathcal{T}^{3}=\mathcal{U}^{2}=(\mathcal{ST})^{3}=(\mathcal{SU})^{2}=(\mathcal{TU})^{2}=(\mathcal{STU})^{4}=I$. Through the standard relation connecting the number of elements with the dimension of the irreducible representations of $S_{4}$---$\mathbf{24=1}^{2}\mathbf{+1}^{2}%
\mathbf{+2}^{2}\mathbf{+3}^{2}\mathbf{+3}^{2}$---we deduce that $S_{4}$ contains five real irreducible representations where we have two singlets $\mathbf{1}$ (trivial) and $\mathbf{1}^{\prime }$, one doublet $\mathbf{2}$ and two triplets $\mathbf{3}$ and $\mathbf{3}^{\prime }$. The character table of $S_{4}$ is shown in Tab. \ref{tab:characters} where the index in $\mathcal{C}_{i}$ indicates the number of elements in each conjugacy class which are given explicitly as follows 
\begin{eqnarray}
\mathcal{C}_{1} &=&\left\{ 1\right\}, \quad \mathcal{C}_{3}=\left\{\mathcal{S},\mathcal{TST}^{2},\mathcal{T}^{2}\mathcal{ST}\right\}, \nonumber \\
\mathcal{C}_{6}&=&\left\{ \mathcal{U},\mathcal{TU},\mathcal{SU},%
\mathcal{T}^{2}\mathcal{U},\mathcal{STSU},\mathcal{ST}^{2}\mathcal{SU}\right\}, \\
\mathcal{C}_{8} &=&\left\{\mathcal{T},\mathcal{ST},\mathcal{TS},\mathcal{%
STS},\mathcal{T}^{2},\mathcal{ST}^{2},\mathcal{T}^{2}\mathcal{S},\mathcal{ST}^{2}\mathcal{S}\right\}, \nonumber \\ 
\mathcal{C}_{6}^{\prime }&=&\left\{\mathcal{STU},\mathcal{TSU},\mathcal{T}^{2}\mathcal{SU},\mathcal{ST}^{2}\mathcal{U},\mathcal{TST}^{2}\mathcal{U},\mathcal{T}^{2}\mathcal{STU}\right\}. \nonumber
\end{eqnarray}
The representation matrices for the $S_{4}$ generators $\mathcal{T}$, $\mathcal{S}$ and $\mathcal{U}$ in the basis where $\mathcal{T}$ is diagonal are of the following form for the five different representations
\begin{widetext}
\begin{eqnarray}
\begin{array}{cccc}
\mathbf{1}_{(1,1,1)}: & \mathcal{S}=1 & \mathcal{T}=1 & \mathcal{U}=1, \\ 
\mathbf{1}_{(1,1,-1)}^{\prime }: & \mathcal{S}=1 & \mathcal{T}=1 & \mathcal{U}=-1, \\ 
\mathbf{2}_{(2,-1,0)}: & \mathcal{S}=\left( 
\begin{array}{cc}
1 & 0 \\ 
0 & 1%
\end{array}%
\right) & \mathcal{T}=\left( 
\begin{array}{cc}
\omega & 0 \\ 
0 & \omega ^{2}%
\end{array}%
\right) & \mathcal{U}=\left( 
\begin{array}{cc}
0 & 1 \\ 
1 & 0%
\end{array}%
\right), \\ 
\mathbf{3}_{(-1,0,-1)}: & \mathcal{S}=\frac{1}{3}\left( 
\begin{array}{ccc}
-1 & 2 & 2 \\ 
2 & -1 & 2 \\ 
2 & 2 & -1%
\end{array}%
\right) & \mathcal{T}=\left( 
\begin{array}{ccc}
1 & 0 & 0 \\ 
0 & \omega ^{2} & 0 \\ 
0 & 0 & \omega%
\end{array}%
\right) & \mathcal{U}=-\left( 
\begin{array}{ccc}
1 & 0 & 0 \\ 
0 & 0 & 1 \\ 
0 & 1 & 0%
\end{array}%
\right), \\ 
\mathbf{3}_{\left( -1,0,1\right) }^{\prime }: & \mathcal{S}=\frac{1}{3}%
\left( 
\begin{array}{ccc}
-1 & 2 & 2 \\ 
2 & -1 & 2 \\ 
2 & 2 & -1%
\end{array}%
\right) & \mathcal{T}=\left( 
\begin{array}{ccc}
1 & 0 & 0 \\ 
0 & \omega ^{2} & 0 \\ 
0 & 0 & \omega%
\end{array}%
\right) & \mathcal{U}=\left( 
\begin{array}{ccc}
1 & 0 & 0 \\ 
0 & 0 & 1 \\ 
0 & 1 & 0%
\end{array}%
\right),%
\end{array}
\label{t2}
\end{eqnarray}
\end{widetext}
where $\omega$ is the cube root of unity: $\omega =e^{2\pi i/3}$. There are $30$ subgroups of $S_{4}$, twenty of which are abelian groups. These are nine $Z_{2}$, four $Z_{3}$, three $Z_{4}$ and four Klein four group $K_{4}\simeq Z_{2}\times
Z_{2}$ generated by
\begin{widetext}
\begin{eqnarray}
Z_{2}&:& \quad \quad \mathcal{S}, \quad \mathcal{TST}^{2}, \quad \mathcal{T}^{2}\mathcal{ST}, \quad \mathcal{U}, \quad \mathcal{SU}, \quad \mathcal{TU}, \quad \mathcal{T}^{2}\mathcal{U},\quad \mathcal{STSU}, \quad \mathcal{ST}^{2}\mathcal{SU}, \nonumber \\ 
Z_{3}&:& \quad \quad \mathcal{T}, \quad \mathcal{ST}, \quad \mathcal{TS}, \quad \mathcal{STS}, \\ 
Z_{4}&:& \quad \quad \mathcal{ST}^{2}\mathcal{U}, \quad \mathcal{TSU}, \quad \mathcal{TST}^{2} \mathcal{U}, \nonumber \\
K_{4}&:& \quad \quad \{\mathcal{S},\mathcal{TST}^{2}\}, \quad \{\mathcal{S},\mathcal{U}\}, \quad \{\mathcal{TST}^{2},\mathcal{T}^{2}\mathcal{U}\}, \quad \{\mathcal{T}^{2}%
\mathcal{ST},\mathcal{TU}\}. \nonumber
\label{eq:S4:subgroups}
\end{eqnarray}    
\end{widetext}
Let us now provide the Kronecker products among the irreducible
representations of the $S_{4}$ group. The products between two
representations are presented bellow
\begin{widetext}
\begin{eqnarray}
\mathbf{1}\otimes \mathbf{R} &=&\mathbf{R\quad },\quad \mathbf{1}^{\prime}\otimes \mathbf{1}^{\prime }=\mathbf{1\quad },\quad \mathbf{1}^{\prime}\otimes \mathbf{2}=\mathbf{2\quad },\quad \mathbf{1}^{\prime }\otimes \mathbf{3}=\mathbf{3}^{\prime }\mathbf{\quad },\quad \mathbf{1}^{\prime}\otimes \mathbf{3}^{\prime }=\mathbf{3}  \notag \\
\mathbf{2}\otimes \mathbf{2} &=&\mathbf{1\oplus 1}^{\prime}\mathbf{\oplus 2\quad },\quad \mathbf{2}\otimes \mathbf{3}=\mathbf{2\otimes 3}^{\prime }=\mathbf{3\oplus 3}^{\prime } \\
\mathbf{3\otimes 3} &=&\mathbf{3}^{\prime }\mathbf{\otimes 3}^{\prime }=\mathbf{1\oplus 2\oplus 3\oplus 3}^{\prime }\mathbf{\quad },\quad \mathbf{3\otimes 3}^{\prime}=\mathbf{1}^{\prime }\mathbf{\oplus 2\oplus 3\oplus 3}^{\prime }  \notag
\end{eqnarray}    
\end{widetext}
where $\mathbf{R}$ stands for any irreducible representation of $S_{4}$. In the following we list the Clebsch-Gordan coefficients using the notations $\alpha \sim \mathbf{1},$ $\alpha ^{\prime }\sim \mathbf{1}^{\prime },$ $(\beta _{1},\beta _{2})^{T},(\gamma _{1},\gamma _{2})^{T}\sim \mathbf{2},$ $%
(a_{1},a_{2},a_{3})^{T},(b_{1},b_{2},b_{3})^{T}\sim \mathbf{3},$ $(c_{1},c_{2},c_{3})^{T},(d_{1},d_{2},d_{3})^{T}\sim \mathbf{3}^{\prime }$.
For a singlet multiplied with a doublet or a triplet we have
\begin{widetext}
\begin{eqnarray}
\mathbf{1}\otimes \mathbf{2} &=& (\alpha \beta_{1},\alpha \beta_{2})^{T}, \quad  \mathbf{1}^{\prime }\otimes \mathbf{2} = (\alpha ^{\prime }\beta_{1},-\alpha^{\prime }\beta_{2})^{T}, \quad \mathbf{1}\otimes \mathbf{3} = (\alpha a_{1},\alpha a_{2},\alpha a_{3})^{T}, \nonumber \\
\mathbf{1}^{\prime }\otimes \mathbf{3} &=& (\alpha^{\prime}a_{1},\alpha^{\prime }a_{2},\alpha^{\prime}a_{3})^{T}, \quad \mathbf{1}\otimes \mathbf{3}^{\prime} = (\alpha c_{1},\alpha c_{2},\alpha c_{3})^{T}, \quad \mathbf{1}^{\prime}\otimes \mathbf{3}^{\prime} = (\alpha^{\prime }c_{1},\alpha^{\prime }c_{2},\alpha^{\prime }c_{3})^{T}.
\end{eqnarray}    
\end{widetext}
For a doublet coupled to a doublet we have
\begin{widetext}
\begin{equation}
\mathbf{2}\otimes \mathbf{2}=\left( 
\begin{array}{c}
\beta _{1} \\ 
\beta _{2}%
\end{array}%
\right) \otimes \left( 
\begin{array}{c}
\gamma _{1} \\ 
\gamma _{2}%
\end{array}%
\right) =\left( \beta _{1}\gamma _{2}+\beta _{2}\gamma _{1}\right) _{\mathbf{%
1}}+\left( \beta _{1}\gamma _{2}-\beta _{2}\gamma _{1}\right) _{\mathbf{1}%
^{\prime }}+\left( 
\begin{array}{c}
\beta _{2}\gamma _{2} \\ 
\beta _{1}\gamma _{1}
\end{array}%
\right) _{\mathbf{2}}
\end{equation}
and for a doublet multiplied with a triplet
\begin{eqnarray}
\mathbf{2}\otimes \mathbf{3} &=&\left( 
\begin{array}{c}
\beta _{1} \\ 
\beta _{2}%
\end{array}%
\right) \otimes \left( 
\begin{array}{c}
a_{1} \\ 
a_{2} \\ 
a_{3}%
\end{array}%
\right) =\left( 
\begin{array}{c}
\beta _{1}a_{2}+\beta _{2}a_{3} \\ 
\beta _{1}a_{3}+\beta _{2}a_{1} \\ 
\beta _{1}a_{1}+\beta _{2}a_{2}%
\end{array}%
\right) _{\mathbf{3}}+\left( 
\begin{array}{c}
\beta _{1}a_{2}-\beta _{2}a_{3} \\ 
\beta _{1}a_{3}-\beta _{2}a_{1} \\ 
\beta _{1}a_{1}-\beta _{2}a_{2}%
\end{array}%
\right) _{\mathbf{3}^{\prime }}  \notag \\
\mathbf{2}\otimes \mathbf{3}^{\prime } &=&\left( 
\begin{array}{c}
\beta _{1} \\ 
\beta _{2}%
\end{array}%
\right) \otimes \left( 
\begin{array}{c}
c_{1} \\ 
c_{2} \\ 
c_{3}%
\end{array}%
\right) =\left( 
\begin{array}{c}
\beta _{1}c_{2}-\beta _{2}c_{3} \\ 
\beta _{1}c_{3}-\beta _{2}c_{1} \\ 
\beta _{1}c_{1}-\beta _{2}c_{2}%
\end{array}%
\right) _{\mathbf{3}}+\left( 
\begin{array}{c}
\beta _{1}c_{2}+\beta _{2}c_{3} \\ 
\beta _{1}c_{3}+\beta _{2}c_{1} \\ 
\beta _{1}c_{1}+\beta _{2}c_{2}%
\end{array}%
\right) _{\mathbf{3}^{\prime }}
\end{eqnarray}%
\end{widetext}
For the products $\mathbf{3\otimes 3}$ and $\mathbf{3}^{\prime }\mathbf{\otimes 3}^{\prime }$ (with $a_{i}$, $b_{i}$ to be replaced by $c_{i}$, $d_{i}$) we get
\begin{widetext}
\begin{eqnarray}
\mathbf{3\otimes 3} &\mathbf{=}&\left( 
\begin{array}{c}
a_{1} \\ 
a_{2} \\ 
a_{3}%
\end{array}
\right) \otimes \left( 
\begin{array}{c}
b_{1} \\ 
b_{2} \\ 
b_{3}%
\end{array}
\right) =\left( a_{1}b_{1}+a_{2}b_{3}+a_{3}b_{2}\right) _{\mathbf{1}}+\left( 
\begin{array}{c}
a_{1}b_{3}+a_{2}b_{2}+a_{3}b_{1} \\ 
a_{1}b_{2}+a_{2}b_{1}+a_{3}b_{3}
\end{array}
\right) _{\mathbf{2}}  \notag \\
&&+\left( 
\begin{array}{c}
a_{2}b_{3}-a_{3}b_{2} \\ 
a_{1}b_{2}-a_{2}b_{1} \\ 
a_{3}b_{1}-a_{1}b_{3}%
\end{array}%
\right) _{\mathbf{3}}+\left( 
\begin{array}{c}
2a_{1}b_{1}-a_{2}b_{3}-a_{3}b_{2} \\ 
2a_{3}b_{3}-a_{1}b_{2}-a_{2}b_{1} \\ 
2a_{2}b_{2}-a_{1}b_{3}-a_{3}b_{1}%
\end{array}%
\right) _{\mathbf{3}^{\prime }}
\end{eqnarray}%
while the Clebsch Gordan coefficients for the product $\mathbf{3\otimes 3}%
^{\prime }$ read
\begin{eqnarray}
\mathbf{3\otimes 3}^{\prime } &=&\left( 
\begin{array}{c}
a_{1} \\ 
a_{2} \\ 
a_{3}%
\end{array}%
\right) \otimes \left( 
\begin{array}{c}
c_{1} \\ 
c_{2} \\ 
c_{3}%
\end{array}%
\right) =\left( a_{1}c_{1}+a_{2}c_{3}+a_{3}c_{2}\right) _{\mathbf{1}^{\prime
}}+\left( 
\begin{array}{c}
a_{1}c_{3}+a_{2}c_{2}+a_{3}c_{1} \\ 
-a_{1}c_{2}-a_{2}c_{1}-a_{3}c_{3}%
\end{array}%
\right) _{\mathbf{2}}  \notag \\
&&+\left( 
\begin{array}{c}
2a_{1}c_{1}-a_{2}c_{3}-a_{3}c_{2} \\ 
2a_{3}c_{3}-a_{1}c_{2}-a_{2}c_{1} \\ 
2a_{2}c_{2}-a_{1}c_{3}-a_{3}c_{1}%
\end{array}%
\right) _{\mathbf{3}}+\left( 
\begin{array}{c}
a_{2}c_{3}-a_{3}c_{2} \\ 
a_{1}c_{2}-a_{2}c_{1} \\ 
a_{3}c_{1}-a_{1}c_{3}%
\end{array}%
\right)_{\mathbf{3}^{\prime }}
\end{eqnarray} 
\end{widetext}

\bibliographystyle{JHEP}
\bibliography{bibliography.bib}

\end{document}